\documentclass[a4paper,11pt]{article}
\pdfoutput=1 

\usepackage{jheppub} 

\usepackage[T1]{fontenc} 
\usepackage{tikz-cd}
\usepackage{caption}
\usepackage{soul}
\usepackage{ulem}

\title{\boldmath  Integrable  Sigma Models  and Universal  Root $T\bar{T}$ Deformation via Courant-Hilbert Approach}

\author[a]{H. Babaei-Aghbolagh,}
\author[a,b,]{Bin Chen,}
\author[a,c]{Song He}

\affiliation[a]{Institute of Fundamental Physics and Quantum Technology, \& School of Physical Science and Technology, Ningbo University, Ningbo, Zhejiang 315211, China}
\affiliation[b]{School of Physics, \&  Center for High Energy Physics, Peking University, No.5 Yiheyuan Rd, Beijing 100871, P. R. China}
\affiliation[c]{Max Planck Institute for Gravitational Physics (Albert Einstein Institute), Am M\"uhlenberg 1, 14476 Golm, Germany}
\emailAdd{hosseinbabaei@nbu.edu.cn}
\emailAdd{chenbin1@nbu.edu.cn}
\emailAdd{hesong@nbu.edu.cn}

\abstract{We develop a unified Courant--Hilbert framework for constructing two-dimensional integrable sigma models deformed by two couplings: a marginal one $\gamma$ and an irrelevant one $\lambda$. The integrability condition is encoded in a nonlinear partial differential equation (PDE) for two invariants $(P_1, P_2)$, whose general solution could be expressed through an arbitrary generating function $\ell(\tau)$. This formulation encompasses and extends known models, such as ModMax and Born-Infeld, while introducing new classes of solvable models with closed-form Lagrangians, including those with logarithmic and $q$-deformations. All resulting theories obey a universal root-$T\overline{T}$ flow equation, consistent under dimensional reduction from four-dimensional duality-invariant electrodynamics. Using perturbative expansions, we recover ModMax in the free limit, determine the $\gamma$-dependence of the coupling functions, and show how different flow equations, including a single-trace form, naturally emerge. Our results reveal deep structural connections between self-duality, integrability, and deformation dynamics across different dimensions. 
}

\begin{document}
\maketitle
\flushbottom
\section{Introduction}\label{0.}

The $T\bar{T}$ operator in two-dimensional quantum field theories presents a remarkable exception to the typical challenges posed by irrelevant operators \cite{Smirnov:2016lqw,Cavaglia:2016oda}, which usually destabilize UV physics and require infinite counterterms \cite{McGough:2016lol}. Unlike generic irrelevant deformations, the $T\bar{T}$ operator is quantum-mechanically well defined and solvable \cite{Cardy:2018sdv}. 
This solvability stems from the operator’s preservation of sufficient symmetries in 2D, enabling the explicit analysis of RG flows that would otherwise be intractable \cite{Cavaglia:2016oda,Rosenhaus:2019utc}, particularly when deforming conformal field theories with such special operators. The solutions of the Thermodynamic Bethe Ansatz equations have been investigated for relativistic theories characterized by factorizable S-matrices in integrable quantum field theories deformed through {  Castillejo–Dalitz–Dyson (CDD) }factors. These S-matrices emerge from generalized $T\bar{T}$ deformations induced by specific irrelevant operators \cite{Camilo:2021gro}.
Consider a quantum field theory described by a seed action 
$S_0$. Introducing an integrated local operator 
$\mathcal{O}_\lambda$, we can formally deform the action through an infinitesimal perturbation:
\begin{equation}
S_\lambda = S_0 + \lambda \, \mathcal{O}_\lambda + \mathcal{O}_\lambda(\lambda^2),
\end{equation}
where $\lambda$ is the deformation parameter, when $S_0$ corresponds to a conformal field theory (CFT) and $\mathcal{O}_\lambda$ represents the $T\bar{T}$-deformation operator. {  Depending on the conformal dimension of the $\lambda$ coupling, the operator $\mathcal{O}_\lambda$ can be either irrelevant or marginal.} This deformation naturally induces an irrelevant flow equation for the deformed theory.  Comprehensive reviews on $T\bar{T}$ deformation are available in \cite{Jiang:2019epa,He:2025ppz}, which analyze the current theoretical framework and its applications.

{ Discussions of a deformation driven by an operator proportional to $\sqrt{T\bar{T}}$, began with reference~\cite{Rodriguez:2021tcz}.}
The root-$T\bar{T}$ deformation~\cite{Babaei-Aghbolagh:2022uij} was first identified in duality-invariant electromagnetic theories to describe the ModMax theory~\cite{Bandos:2020jsw},  and later extended to two-
dimensional
bosonic systems \cite{Babaei-Aghbolagh:2022leo,Conti:2022egv,Ferko:2022cix,Ebert:2024zwv}, integrable models 
\cite{Borsato:2022tmu,Ferko:2024ali,Cesaro:2024ipq,Cesaro:2025msv,Ferko:2025bhv}, chiral boson theories \cite{Ebert:2024zwv}, { Hamiltonian analysis \cite{Tempo:2022ndz}} 
higher-spin theories \cite{Bielli:2024ach,Bielli:2025uiv}, 
holography~\cite{Ebert:2023tih,Ran:2025xas,tian2024modulartransformationsonshellactions,Liang:2025vmx,Garcia:2022wad}, p-form gauge 
 theories~\cite{Babaei-Aghbolagh:2020kjg,Babaei-Aghbolagh:2022itg,Ferko:2024zth,Kuzenko:2025jgk,Hutomo:2025dfx}, partition functions~\cite{He:2025fdz}, and emergent gravity \cite{Babaei-Aghbolagh:2024hti,Li:2025lpa,Adami:2025pqr}. Unlike the well-known $T\bar{T}$
deformation (which is an irrelevant operator), the root-$T\bar{T}$ operator is a marginal deformation \cite{Babaei-Aghbolagh:2022leo}.  The root-$T\bar{T}$ deformation represents a novel class of integrable quantum field theories in two dimensions.
Given an original QFT by seed Lagrangian $\mathcal{L}_0$, the  root-$T\bar{T}$ and  irrelevant deformations  are defined by the two flow equations in arbitrary dimensions \cite{Babaei-Aghbolagh:2022uij,Babaei-Aghbolagh:2022leo,Conti:2022egv,Ferko:2022cix,Babaei-Aghbolagh:2024uqp}:
\begin{eqnarray}\label{Rtoglan}
	&&\frac{\partial \mathcal{L}(\gamma,\lambda)}{\partial \gamma} =\frac{1}{\sqrt{d}}\, \sqrt{T(\gamma,\lambda)_{\mu\nu}T(\gamma,\lambda)^{\mu\nu}-\frac{1}{d} {T(\gamma,\lambda)_{\mu}}^{\mu} {T(\gamma,\lambda)_{\nu}}^{\nu}}\,;\\
	&&\frac{\partial \mathcal{L}(\gamma,\lambda)}{\partial \lambda} =\sum_{m} C_m   \Big(T(\gamma,\lambda)_{\mu\nu}T(\gamma,\lambda)^{\mu\nu}\Big)^{1-\tfrac{m}{2}} \,\,\Big({T(\gamma,\lambda)_{\mu}}^{\mu} {T(\gamma,\lambda)_{\nu}}^{\nu}\Big)^{\tfrac{m}{2}}\,\,.\nonumber
\end{eqnarray}
where $T(\gamma,\lambda)_{\mu\nu}$ is the stress-energy tensor of the deformed theory with two coupling constants of $\gamma$ and $\lambda$.  And $C_m$ are the constant coefficients of the deformed theory, where $m =0, 1, 2, \dots\,$.  The root-$T\bar{T}$ operator commutes with the irrelevant $T\bar{T}$ operator~\cite{Babaei-Aghbolagh:2022leo}. The solvability of these theories stems from the root-$T\bar{T}$ as a composite operator constructed from the components of the Lax connection in 2D integrable QFTs~\cite{Conti:2018jho,Sakamoto:2025hwi}. These results establish root-$T\bar{T}$ as a bridge between 4D duality-symmetric theories and 2D integrable models~\cite{Kuzenko:2013pha,Kuzenko:2023ysh,Ferko:2023pr,Ferko:2022iru,Babaei-Aghbolagh:2025uoz}.
The root-$T\bar{T}$ deformation has emerged as a powerful tool for constructing and analyzing integrable quantum field theories, with deep connections to Lax formalism and classical integrability. As demonstrated in \cite{Borsato:2022tmu}, this deformation preserves the integrable structure of 2D field theories by maintaining an infinite tower of conserved charges, a hallmark of integrability. Ref.~\cite{Borsato:2022tmu} has shown that root-$T\bar{T}$ flows can be understood as particular transformations of the Lax connection that leave the zero-curvature condition invariant, thereby ensuring the survival of integrability under deformation.
 According to~\cite{Conti:2018jho, Chen:2021aid}, the deformation induces a specific modification of the Lax pair that preserves the spectral parameter dependence while altering the Hamiltonian structure. This property allows root-$T\bar{T}$ to generate families of integrable models from seed theories while maintaining solvability. The relationship between root-$T\bar{T}$ and Lax connections becomes particularly significant when examining classical integrable systems, so that the integrability condition is reduced to a partial differential equation (PDE) and the solutions to the PDE are deformed theories with energy-momentum tensors \cite{Babaei-Aghbolagh:2024uqp, Babaei-Aghbolagh:2025lko, Babaei-Aghbolagh:2025cni}. 

 The primary motivation of this paper is to identify integrable theories that satisfy the given PDE. The new integrable theories we present here, introduced in this work, extend beyond the 2d counterparts of Born-Infeld-like and ModMax-like theories. These two-dimensional integrable sigma models possess closed-form solutions and encompass a broad class of $\mathbf{q}$-deformed and logarithmic theories. 
We employ the Courant-Hilbert approach to solve the PDE, which yields novel integrable systems. Remarkably, the two-dimensional integrable theories obtained through this method are consistent with the dimensional reduction of four-dimensional electrodynamics, as recently discovered by Russo and Townsend. These theories feature two distinct couplings, $\gamma$ and $\lambda$, which follow a root flow equation and an irrelevant flow equation.

The second motivation of this paper is to investigate the root flow equation for the newly obtained integrable theories. We demonstrate that all of these integrable theories obey the same root flow equation. Furthermore, we establish that this flow equation represents a universal equation governing a broad class of two-dimensional integrable models. 
Inspired by the four-dimensional root flow equation, we reveal a unified formulation of the root flow equation that applies to both four-dimensional and two-dimensional theories. This universal representation demonstrates the connection between four-dimensional electrodynamics and two-dimensional integrable theories.

The organization of this paper is as follows. In Section~\eqref{1.}, we derive the integrability condition as a partial differential equation for sigma models with two couplings. In Section~\eqref{30}, we solve this PDE using the Courant-Hilbert approach, establishing a systematic method to generate integrable models. Section~\eqref{261} applies this method to construct a rich class of theories, including generalized Born-Infeld-like, $q$-deformed, and logarithmic models---all governed by two distinct flow parameters, $\gamma$ and $\lambda$. In Section~\eqref{05}, we derive a universal representation of the root $T\bar{T}$ operator and prove that all models constructed obey a common flow equation, invariant under dimensional reduction from 4D to 2D. 
in Section~\eqref{ff8412ff2}, using a perturbative expansion of the characteristic function, we derive the Lagrangian up to $\mathcal{O}(\lambda^3)$ and determine the specific $\gamma$-dependence of the coupling functions $f_i(\gamma)$ required to satisfy a general root flow equation. We demonstrate that for the special case $e_2 = 1/4$, the flow equation simplifies to $\partial_\gamma \mathcal{L} = \tau \dot{\ell}(\tau)$, { where  $\dot{\ell}(\tau) = d\ell(\tau) /d\tau$ and $\ell(\tau)$ is a real
function of $\tau$ called Courant-Hilbert function } and show how other flow equations, including a single-trace form, arise from different choices of the deformation.
We conclude in Section~\eqref{06kk} with a discussion of potential generalizations and implications for a non-conformal family of  ModMax theories and string-theoretic frameworks.

\section{Integrability condition} \label{1.}
The study of integrable structures in the worldsheet formulation of string theory has motivated the search for integrability-preserving transformations in two-dimensional integrable quantum field theories. In particular, identifying integrable deformations of 2D sigma models is of significant interest for applications in string theory.

We focus on classical field theories defined on a flat, two-dimensional spacetime manifold, denoted as $\Sigma$. Occasionally, we refer to $\Sigma$ as the worldsheet and choose coordinates $\sigma^{\alpha} = (\sigma, \tau)$ on $\Sigma$.

Specifically, we consider two-dimensional sigma models whose target space $G$ is a Lie group with its Lie algebra denoted as $\mathfrak{g}$. The fundamental field of these models, represented as $g(\sigma, \tau)$, maps the worldsheet $\Sigma$ into $G$. From this field $g$, two crucial quantities can be constructed: the left- and right-invariant Maurer-Cartan forms, defined as follows:
\begin{equation} \label{P1P21}
	j = g^{-1} dg, \quad \quad \quad \tilde{j} = -(dg) g^{-1} \,.
\end{equation}
Both $j$ and $\tilde{j}$ satisfy the flatness condition, which can be expressed using light-cone coordinates\footnote{For any vector $A^\mu$, we define $A^\pm = \tfrac{1}{2}(A^0 \pm A^1)$.}:
\begin{equation} \label{jj}
	\partial_+ j_- - \partial_- j_+ + \left[j_+, j_-\right] = 0 = \partial_+ \tilde{j}_- - \partial_- \tilde{j}_+ + \left[\tilde{j}_+, \tilde{j}_-\right] \,.
\end{equation}

One of the simplest examples of such a sigma model is the Principal Chiral Model (PCM). The Lagrangian of the PCM can be expressed in terms of $j$ or $\tilde{j}$:
\begin{equation} \label{eq:Lpcm}
	\mathcal{L}_{PCM} = \frac{1}{2} g^{\mu\nu} \text{tr}\left[j_\mu j_\nu\right] = -\frac{1}{2} \text{tr}\left[j_+ j_-\right] \,.
\end{equation}
{ where $g^{\mu\nu} $ is the two dimensional inverse-metric.} An equivalent formulation is given by
\begin{equation}
	\mathcal{L}_{PCM} = \frac{1}{2} g^{\mu\nu} \text{tr}\left[\tilde{j}_\mu \tilde{j}_\nu\right] = -\frac{1}{2} \text{tr}\left[\tilde{j}_+ \tilde{j}_-\right] \,.
\end{equation}

Any Lagrangian that depends on $j_{\pm}$ only through the combinations of $\,\text{tr}\,[j_+ j_-]\,$ and $\text{tr}(j_+ j_+)  \text{tr}[j_- j_-]$ can be rewritten, after a change of variables, as a function $\mathcal{L}(P_1, P_2)$ of the two variables:
\begin{equation} \label{P1P2}
	P_1 = -\text{tr}[j_+ j_-], \quad P_2 = \frac{1}{2} \left(\text{tr}[j_+ j_+] \text{tr}[j_- j_-] + (\text{tr}[j_+ j_-])^2 \right) \,.
\end{equation}
The equation of motion for the Lagrangian $\mathcal{L}(P_1, P_2)$ can be expressed as $\partial_\mu \mathfrak{J}^\mu = 0$, where $\mathfrak{J}^\mu$ is the Noether current associated with the symmetry under right-multiplication of $g$ by a general group element:
\begin{equation} \label{eq:deformedPCMeomgeneric}
	\mathfrak{J}_\mu = 2 \frac{\partial \mathcal{L}(P_1, P_2)}{\partial P_1} j_\mu + 4 \frac{\partial \mathcal{L}(P_1, P_2)}{\partial P_2} g^{\nu\rho} \text{tr}[j_\mu j_\nu] j_\rho \,.
\end{equation}

For  the deformed theories $\mathcal{L}^{(\lambda,\gamma)}(P_1, P_2)$, characterized by the couplings $\gamma$ and $\lambda$ and dependent on the independent variables $P_1$ and $P_2$, their equations of motion are equivalent to the flatness condition of the Lax connection:
\begin{equation} \label{flatness}
	0 = \partial_{+}\mathfrak{L}^{(\lambda,\gamma)}_{-} - \partial_{-}\mathfrak{L}^{(\lambda,\gamma)}_{+} + [\mathfrak{L}^{(\lambda,\gamma)}_+, \mathfrak{L}^{(\lambda,\gamma)}_-] \,,
\end{equation}
where the Lax connection takes the form
\begin{equation}
	\mathfrak{L}_{\pm}^{(\lambda,\gamma)} = \frac{j_\pm \pm z \mathfrak{J}_{\pm}}{1 - z^2} \,.
\end{equation}
This flatness condition holds when the current $\mathfrak{J}_{\alpha}$ satisfies the relation $\big[\mathfrak{J}_+, j_-\big] = \big[j_+, \mathfrak{J}_-\big]$. As demonstrated in Ref.~\cite{Borsato:2022tmu,Babaei-Aghbolagh:2025lko}, this equality is valid if and only if the following differential equation is satisfied:
\begin{equation} \label{PDE2D}
	8 (P_1^2 - P_2) \left(\frac{\partial \mathcal{L}^{(\lambda,\gamma)}}{\partial P_2}\right)^2 + 8 P_1 \frac{\partial \mathcal{L}^{(\lambda,\gamma)}}{\partial P_2} \frac{\partial \mathcal{L}^{(\lambda,\gamma)}}{\partial P_1} + 4 \left(\frac{\partial \mathcal{L}^{(\lambda,\gamma)}}{\partial P_1}\right)^2 = 1 \,.
\end{equation}
Recently, the integrability condition Eq.~\eqref{PDE2D} has been identified as both a necessary and sufficient condition for $O(d, d)$ duality in two-dimensional scalar theories \cite{Babaei-Aghbolagh:2025lko}. These two-dimensional scalar theories are closely related to the integrable theories examined in this paper. { It was shown in \cite{PhysRevLett.134.101601} that the T-duality transformation commutes with the stress-energy tensor transformation.}

\section{General solution of the  integrability condition by using Courant-Hilbert function }\label{30}
In this section, motivated by the theoretical framework of self-dual nonlinear electrodynamics in four dimensions, we systematically reduce the integrability condition for two-dimensional scalar theories, expressed as the (PDE) ~\eqref{PDE2D}, to a simpler form through an appropriate change of variables. This reduction builds upon the classical approach developed by Courant and Hilbert, which introduces an auxiliary field to solve the simplified differential equation without introducing additional dynamical degrees of freedom to the theory. 

Applying this methodology to the integrability condition, we derive novel integrable theories. Two distinct coupling parameters characterize the resulting theories:
\begin{itemize}
    \item $\gamma$: a dimensionless coupling constant that corresponds to a marginal flow equation
    \item $\lambda$: a dimensionful coupling constant that corresponds to an irrelevant flow equation
\end{itemize}
This approach preserves the integrability structure of the original theory and reveals a richer class of exactly solvable models with modified interaction properties.
\subsection{Courant-Hilbert solution for duality invariant condition in (NED) theories}\label{301}
The integrability condition in two dimensions, represented by the differential equation ~\eqref{PDE2D}, is closely related to the duality-invariant condition for nonlinear electromagnetic field theories in four dimensions. In four dimensions, the duality invariant condition is expressed as a differential equation~\cite{Bialynicki-Birula:1984daz, Bialynicki-Birula:1992rcm, Gaillard:1981rj, Gaillard:1997rt, Gibbons:1995cv, Gibbons:1995ap,Chen:2025ndc,Murcia:2025psi}:  
\begin{eqnarray}
	\label{lagrangeNGZ4}
	\Big (\frac{\partial \mathcal{L}}{\partial S}\Big )^2\, -2 \frac{S}{P} \,\frac{\partial \mathcal{L}}{\partial S} \frac{\partial \mathcal{L}}{\partial P}-\Big (\frac{\partial \mathcal{L}}{\partial P}\Big )^2\, =1\,,
\end{eqnarray}
where $S=-\frac{1}{4}F_{\mu\nu}F^{\mu\nu}$ and $\,\,P=-\frac{1}{4}F_{\mu\nu}\tilde F^{\mu\nu}$ are two Lorentz invariant variables. By adjusting $ U = \tfrac{1}{2}(\sqrt{S^2 + P^2} - S) $ and $ V = \tfrac{1}{2}(\sqrt{S^2 + P^2} + S) $, differential equation ~\eqref{lagrangeNGZ4} can be expressed as follows~\cite{Gaillard:1997rt}:
\begin{eqnarray}
	\label{dUdN4}
	\frac{\partial \mathcal{L}}{\partial U} \frac{\partial \mathcal{L}}{\partial V}=-1\,.
\end{eqnarray}
The general solution to this equation was first provided by Courant and Hilbert, as referenced in \cite{Courant:Hilbert}.The general solution is implicitly expressed in terms of $(U, V)$ and a real function $\ell (\tau)$ of an auxiliary field $\tau$ as follows:
\begin{eqnarray}
	\label{SPDE4}
	\mathcal{L}=\ell (\tau)-\frac{2U}{\dot\ell(\tau)}\ , 
	\qquad  \tau=V+\frac{U}{\dot\ell^2(\tau)}\, . 	
\end{eqnarray}
We will refer to this as the Courant-Hilbert (CH) solution and denote $\ell (\tau)$ as a 'CH-function'. Recently, Russo and Townsend have explored the solution of the differential equation in ~\eqref{dUdN4} for four-dimensional electromagnetic theories with two couplings, $\lambda$, and $\gamma$, using the Courant-Hilbert approach in  \cite{Russo:2024xnh, Russo:2024dual}.
Additionally, a general perturbation solution for the differential equation ~\eqref{lagrangeNGZ4} is presented in Ref. \cite{Babaei-Aghbolagh:2024uqp}, which includes the solutions of Russo and Townsend in \cite{Russo:2024csd}.
\subsection{Courant-Hilbert solution for integrable condition }\label{301}
In this section, we apply the Courant-Hilbert approach to solving the two-dimensional partial differential equation ~\eqref{PDE2D}, presenting a general solution to identify integrable theories.
We demonstrate that integrable theories, such as the Principal Chiral Model and the two-dimensional Maximal Model, can be derived using the Courant-Hilbert approach. To solve the differential equation  ~\eqref{PDE2D}, we consider the following variable:  
\begin{eqnarray}
	\label{UN}
	q_1=\frac{ 1}{4} (\sqrt{- {P_{1}}^2 + 2 P_{2}{}} -P_1)\, ,\quad q_2=\frac{ 1}{4}(\sqrt{- {P_{1}}^2 + 2 P_{2}{}} +P_1)\, .
\end{eqnarray}
Note that both $q_1$ and $q_2$ are non-negative. The  PDE ~\eqref{PDE2D} can be reformulated using the variables $q_1$ and $q_2$ as follows:
\begin{eqnarray}
	\label{dUdN}
	\frac{\partial \mathcal{L}^{(\lambda,\gamma)}}{\partial q_1} \,\,\, \frac{\partial \mathcal{L}^{(\lambda,\gamma)}}{\partial q_2} =-1 \,.
\end{eqnarray}
The general solution to this equation in the positive $(q_1, q_2)$-quadrant, given the initial conditions $\mathcal{L}(0, q_2) = \ell(q_2)$, is
\begin{equation}
	\label{gensol}
	\mathcal{L}=\ell (\tau)-\frac{2\, q_1}{[\dot {\ell }(\tau)]}\ ,\qquad 
	\tau = q_2 +\frac{q_1}{ \,\, [ \dot \ell (\tau)]^2}\, , 
\end{equation}
where $\ell (\tau)$ is an arbitrary real function  of  the real variable $\tau$. It follows from equation ~\eqref{gensol} that 
$\tfrac{\partial \mathcal{L}}{\partial q_1}=-\tfrac{1}{\dot {\ell }(\tau)} $ and $\tfrac{\partial \mathcal{L}}{\partial q_2}=\dot {\ell }(\tau) $, and consequently $\tfrac{\partial \mathcal{L}}{\partial q_1} \,\,\tfrac{\partial \mathcal{L}}{\partial q_2}=-1$.

As illustrations, let us derive the integrable theories discussed in \cite{Borsato:2022tmu} by using Courant-Hilbert solutions. These theories include the free theory without coupling, known as the Principal Chiral Model (PCM), a theory with one coupling derived from the root $T\bar{T}$-deformation of the PCM, referred to as the two-dimensional ModMax-like theory.
\begin{itemize}
	\item 
	Principal Chiral Model:\\
	A straightforward choice for the CH function in CH solutions is $\ell (\tau)= \tau$. With this choice, we obtain $\dot \ell (\tau)=1$, which, when substituted into ~\eqref{gensol}, results in:
	\begin{equation}\
		\mathcal{L}_{PCM}=- q_1 +   q_2 \,, \qquad  \qquad \tau= \,q_1  + q_2 . 
	\end{equation}
	By utilizing Eqs. ~\eqref{UN} and ~\eqref{P1P2}, we can express the Lagrangian of the Principal Chiral Model in terms of $j^\alpha$ as follows:
	\begin{equation}\label{LPCM}
		\mathcal{L}_{PCM}=\frac{1}{2}  P_1=-\frac{1}{2}\text{tr}\left[j_+\,j_-\right] . 
	\end{equation}
	The above Lagrangian represents the simplest integrable theory derived from the CH approach.	
\end{itemize}
\begin{itemize}
	\item 
	Two-dimensional ModMax{-like}  theory:\\
	The two-dimensional ModMax{-like}  theory is a marginally deformed version of PCM. Consequently, the CH function of this theory should be linear with respect to $\tau$. Thus, we consider the CH function of this theory as $\ell (\tau)= e^{\gamma} \tau$, scaled by the coupling constant $\gamma$. With this choice, we have $\dot \ell = e^{\gamma} $. Consequently, it can be shown that $\ell =  \tau \dot \ell $  is valid. As shown in \cite{Russo:2024xnh,Russo:2024dual}, the condition $\ell = \tau \dot{\ell}$ establishes the traceless condition for ModMax theories.
	This CH function yields: 
	\begin{equation}\label{LMM}
		\mathcal{L}_{MM}=-e^{-\gamma} q_1 +  e^{\gamma} q_2 \,, \qquad  \qquad  \tau= e^{-2 \gamma}\,q_1  + q_2 . 
	\end{equation}
	Using Eq. ~\eqref{UN}, we can express the Lagrangian ~\eqref{LMM} in terms of the two variables $P_1$ and $P_2$ as follows:
	\begin{equation}\label{LMM1}
		\mathcal{L}_{MM}=\frac{1}{2} \biggl(\cosh(\gamma) P_1 + \sqrt{- P_1^2 + 2 P_2} \sinh(\gamma)\biggr). 
	\end{equation}
	From Eq. ~\eqref{P1P2}, we have $-P_1^2+2P_2=\text{tr}[j_+ j_+]\text{tr}[j_- j_-]$, and thus the Lagrangian of 2D ModMax{-like}  will be as follows:
	\begin{equation}\label{LMM2}
		\mathcal{L}_{MM}=\frac{1}{2}\left(-\cosh(\gamma)\text{tr}[j_+j_-]+\sinh(\gamma)\sqrt{\text{tr}[j_+j_+]\text{tr}[j_-j_-]}\right).
	\end{equation}
\end{itemize}
The 2D ModMax{-like}  Lagrangian, presented in Eq.~\eqref{LMM2}, is an integrable theory derived from the Courant-Hilbert (CH) approach, with its integrability thoroughly examined in \cite{Borsato:2022tmu}. In the limit of $\gamma \to 0$, Lagrangian ~\eqref{LMM2} simplifies to the Lagrangian of the Principal Chiral Model (PCM) in  ~\eqref{LPCM}.
\section{Integrable models with two coupling constants from CH solutions}\label{261}
In this section, we extend beyond the well-known models to demonstrate that integrable models with two coupling constants, $\gamma$ and $\lambda$, such as the general Born-Infeld theory in 2D, can be derived using the Courant-Hilbert approach. The two-dimensional general Born-Infeld-like theory, characterized by these coupling constants, includes two flow equations: marginal and irrelevant, corresponding to $\gamma$ and $\lambda$, respectively. We show that different choices of the Courant-Hilbert function can generate all these theories in CH solutions. Additionally, we reveal that some new theories encompass logarithmic and q-deformed integrable theories.
The CH function can incorporate higher orders of $\tau$ with a dimensionful coupling constant of  $\lambda$. One notable property of the function $\ell(\tau)$ is its power series expansion around $\tau = 0$, ensuring that:
\begin{equation}\label{expan}
	\ell (\tau) = e^\gamma \tau + \lambda \, O(\tau^2) \, . 
\end{equation}
All theories with double coupling constants  $\gamma$ and $\lambda$, derived from CH functions with the property described in ~\eqref{expan}, reduce to the 2D ModMax-like theory in the limit $\lambda  \to 0$. Additionally, in the context of $\gamma \to  0$, these theories are an irrelevant deformation of the Principal Chiral Model (PCM).
\subsection{Two-dimensional general Born-Infeld{-like} theory from CH solutions}\label{26}
The general Born-Infeld theory is an irrelevant $T\bar{T}$ deformation of ModMax theory, specifically with respect to the $\lambda$ coupling constant.  By choosing a suitable CH function with an expansion similar to equation ~\eqref{expan}, we expect to derive the two-dimensional integrable general Born-Infeld theory.
To achieve this goal, the appropriate choice is  
\begin{eqnarray}\label{lBI}
	\ell (\tau) = - \frac{1}{\lambda} \bigg(1 - \sqrt{1 + 2 \lambda \, e^{\gamma} \, \tau}\bigg)\,,
\end{eqnarray}
which represents an expansion of the CH function in the form of $\ell (\tau) =e^\gamma \,  \tau - \frac{1}{2} \, \lambda \,  e^{2\gamma} \, \tau^2 + \frac{1}{2} \, \lambda^2 \, e^{3\gamma} \, \tau^3 $ and corresponds to ~\eqref{expan}.
By utilizing this CH function, we obtain the Lagrangian of the Born-Infeld-like theory as follows:
\begin{eqnarray}\label{Gq1q2}
	\mathcal{L}_{GBI} =\frac{1}{\lambda} \Big(-1 + \sqrt{1 - 2 e^{-\gamma} \lambda q_1} \sqrt{1 + 2 e^{\gamma} \lambda q_2}\Big) \,, \qquad  \tau=\frac{e^{-\gamma} q_1 + e^{ \gamma} q_2}{e^{\gamma} - 2 \lambda q_1}.  
\end{eqnarray}
The expansion of the generalized Born-Infeld{-like} Lagrangian around $\lambda = 0$ is given by
\begin{eqnarray}\label{Gp1p22}
	\mathcal{L}_{GBI} =- e^{-\gamma} q_1 + e^{\gamma} q_2 -  \tfrac{1}{2} e^{-2 \gamma} \lambda (q_1 + e^{2 \gamma} q_2)^2 + \tfrac{1}{2} e^{-3 \gamma} \lambda^2 (- q_1 + e^{2 \gamma} q_2) (q_1 + e^{2 \gamma} q_2)^2.  
\end{eqnarray}
In the weak-field limit at $\lambda=0$, the expansion of the Born-Infeld-like Lagrangian in ~\eqref{Gp1p22},  reduces to the ModMax theory as described in Eq. ~\eqref{LMM}.
By substituting ~\eqref{UN} in the above Lagrangian, we obtain the two-dimensional integrable Born-Infeld-like theory as follows:
\begin{eqnarray}\label{GMSP}
	\mathcal{L}_{GBI} =\frac{1}{ \lambda}\Big(-1 + \sqrt{1+  \lambda \, \biggl(\cosh(\gamma) P_1 + \sqrt{- P_1^2 + 2 P_2} \sinh(\gamma)\biggr)+ \tfrac{1}{2} \lambda^2 (P_1^2 -  P_2) }\Big)\,.
\end{eqnarray}
It can be explicitly verified that the Lagrangians ~\eqref{Gq1q2} and ~\eqref{GMSP} are solutions to the differential equations ~\eqref{dUdN} and ~\eqref{PDE2D}, respectively. Consequently, the Noether current of the two-dimensional general Born-Infeld{-like} theory satisfies the property $\big[\mathfrak{J}+, j-\big] = \big[j_+, \mathfrak{J}_-\big] $, indicating that the theory is integrable.

So far, we have successfully developed integrable theories for PCM ~\eqref{LPCM}, ModMax{-like}  ~\eqref{LMM2}, and general Born-Infeld{-like} ~\eqref{GMSP} using the CH approach, aligning with the findings of Ref. \cite{Borsato:2022tmu}. We aim to demonstrate that the CH approach is a comprehensive method capable of producing integrable theories beyond those mentioned above.
\subsection{Integrable $\mathbf{q}$-deformed  theories}\label{25a}
{  Inspired by the $\mathbf{q}$-deformed electromagnetic theories \cite{Russo:2024csd}, we construct new two-dimensional integrable models using the CH approach.} This section examines a broad class of CH functions that yield a diverse range of integrable theories. These functions are $\mathbf{q}$-deformed CH functions.
We introduce the $\mathbf{q}$-deformed CH functions as follows:
\begin{eqnarray}\label{qDef}
	\ell (\tau) =\frac{1}{\lambda} \Big(1 -  (1 - \tfrac{1}{\mathbf{q}} e^{\gamma} \lambda \tau )^{\mathbf{q}}\Big)\,. 
\end{eqnarray}
The expansion of the $\mathbf{q}$-deformed CH functions on the $\lambda = 0$ is given by
\begin{equation}\label{qD}
	\ell (\tau)=e^{\gamma} \tau -  \frac{e^{2 \gamma} (\mathbf{q}-1)}{2  \mathbf{q}}  \lambda \tau^2+ \frac{e^{3 \gamma} (\mathbf{q}-2) (\mathbf{q}-1) }{6  \mathbf{q}^2}\lambda^2 \tau^3  -  \frac{e^{4 \gamma} (\mathbf{q}-3) (\mathbf{q}-2) (\mathbf{q}-1)}{24 \mathbf{q}^3}  \lambda^3 \tau^4+... \,.
\end{equation}
Although it is not possible to explicitly solve equation $\tau = q_2 +\frac{q_1}{ \,\, [ \dot \ell (\tau)]^2}\, $ with the $\mathbf{q}$-deformed CH functions in ~\eqref{qDef}, these $\mathbf{q}$-deformed CH functions reproduce a wide range of theories. $\mathbf{q}$-deformed theories not only cover the findings of the previous sections but also generate new, potentially significant theories. As anticipated, the $\mathbf{q}$-deformed CH function in Eq.~\eqref{qDef} simplifies to the CH function of the ModMax-like theory when $\mathbf{q} \to 1$ or $\lambda \to  0$. Additionally, when $\mathbf{q} = 1/2$, this function represents the  CH function of the general Born-Infeld-like theory. \footnote{For $\mathbf{q} = \frac{1}{2}$, to align with the results from Sec. ~\eqref{26}, we need to use $\lambda \to - \lambda$. We selected the CH function as ~\eqref{lBI} to compare our findings with those in \cite{Borsato:2022tmu}.} In the remainder of this section, we will explore $\mathbf{q}$-deformed theories with $\mathbf{q} = 3/2$ and $\mathbf{q} = 3/4$, both of which have exact solutions. Additionally, we will investigate the limit of the $\mathbf{q}$-deformed theory as $\mathbf{q} \to \infty$.
\subsubsection{$\mathbf{q}=\frac{3}{2}$ model}\label{2.3}
The Lagrangian  densities can also be found explicitly in the $\mathbf{q}=\frac{3}{2}$  model, where
\begin{eqnarray}\label{q32}
	\ell (\tau) =\frac{1}{\lambda} \Big(1 -  (1 -  \tfrac{2}{3} e^{\gamma} \lambda \tau )^{\tfrac{3}{2}}\Big).
\end{eqnarray}
Using ~\eqref{gensol}, we find
\begin{eqnarray}\label{Noq1q2}
	\tau=\frac{3\, e^{-\gamma}}{2\lambda} -  \frac{e^{-\tfrac{3}{2} \gamma} }{4 \lambda}\Gamma\,,
\end{eqnarray}
where
\begin{eqnarray}\label{NoMg}
	\Gamma = e^{\tfrac{1}{2} \gamma} (3 - 2 e^{\gamma} \lambda q_2) +\sqrt{-24\, \lambda \, q_1 + e^{\gamma} (3 - 2 e^{\gamma} \lambda q_2)^2}\,.
\end{eqnarray}
By applying the relations ~\eqref{gensol} and ~\eqref{Noq1q2}, we derive the Lagrangian for the $\mathbf{q}=\frac{3}{2}$ model:
\begin{eqnarray}\label{NoMq}
	\mathcal{L}_{\mathbf{q}=\tfrac{3}{2}} =\frac{1}{\lambda}+ \frac{e^{- \tfrac{1}{4} \gamma}  (-3 + 2 \lambda e^{\gamma} q_2)}{\sqrt{6}\lambda}\, \Gamma^{\tfrac{1}{2}} + \frac{e^{-\tfrac{3}{4} \gamma}}{3 \sqrt{6} \lambda}  \,\Gamma^{\tfrac{3}{2}}\,.
\end{eqnarray}
It can be explicitly shown that Lagrangian ~\eqref{NoMq} holds within differential equation ~\eqref{dUdN}, and the solution to this differential equation verifies its validity. As a result, the $\mathbf{q}$-deformed theory at $\mathbf{q}=\frac{3}{2}$ is an integrable model derived from the CH approach. The $\mathbf{q}$-deformed theory at $\mathbf{q}=\frac{3}{2}$ aligns with the reduction of the four-dimensional nonlinear causal duality invariant electrodynamic theory, referred to as the no maximum-$\tau$ theory in Ref. \cite{Russo:2024csd}.
\subsubsection{$\mathbf{q}=\frac{3}{4}$ model}\label{422}
Model $\mathbf{q}=\frac{3}{4}$, characterized by the CH function, 
\begin{eqnarray}\label{lDe34}
	\ell (\tau) =\frac{1}{\lambda} \Big(1 -  (1 -  \tfrac{4}{3} e^{\gamma} \lambda \tau )^{\tfrac{3}{4}}\Big)\,,
\end{eqnarray}
is another instance that can be exactly computed within $\mathbf{q}$-deformed theories. Apply Eq. ~\eqref{gensol}, we can find that
\begin{eqnarray}\label{qDe34}
	\tau=\frac{1}{3}e^{-4 \gamma} \, \big(\Delta - 2\, e^{\gamma} \lambda \, {q_1}^2 + q_{2} \big)\,\,,\,\,\,\,\,\,\,\,\,\,\,\,\,\, \Delta=\sqrt{e^{2 \gamma} q_1^2 (9 e^{2 \gamma} + 4 \lambda^2 q_1^2 - 12 e^{3 \gamma} \lambda q_2)}\,.
\end{eqnarray}
The corresponding Lagrangian density is  
\begin{eqnarray}\label{qdef}
	\mathcal{L}_{\mathbf{q}=\tfrac{3}{4}}& =&\frac{1}{\lambda} - 2 e^{-\gamma} q_1 \biggl(1 -  \tfrac{4}{9} e^{-3 \gamma} \lambda (\Delta - 2 e^{\gamma} \lambda q_1^2 + 3 e^{4 \gamma} q_2)\biggr)^{1/4}\\
	&& -  \frac{1}{\lambda} \biggl(1 -  \tfrac{4}{9} e^{-3 \gamma} \lambda (\Delta - 2 e^{\gamma} \lambda q_1^2 + 3 e^{4 \gamma} q_2)\biggr)^{3/4}\nonumber\,.
\end{eqnarray}
Like the other Lagrangians discussed in this paper, it can be shown that the Lagrangian in Eq.~\eqref{qdef} satisfies the differential equation given in Eq.~\eqref{dUdN} and corresponds to an integrable theory. Additionally, $\mathbf{q}=\frac{3}{4}$ model aligns with the duality invariant electromagnetic theory reduction proposed in \cite{Russo:2024dual}.
\subsubsection{$\mathbf{q}$-deformed  theories at  $ \mathbf{q} \to \infty$}\label{263}
It is known that the q-deformed CH function in Eq. ~\eqref{qDef} approaches the CH function of ModMax{ -like}  theory as $ \mathbf{q} \to 0 $. We can set the upper limit for this CH function as $\mathbf{q} \to \infty$. By considering the limit $\mathbf{q} \to \infty$ for $\mathbf{q}$-deformed CH function, we obtain:
\begin{eqnarray}\label{qbi}
	\ell (\tau) = \frac{1}{\lambda} \Big( 1 -  e^{- e^{\gamma} \lambda \tau} \Big) \,,
\end{eqnarray}
The expansion of this CH function in terms of $\lambda^n$ is $\ell (\tau) = e^{\gamma} \tau -  \tfrac{1}{2} e^{2 \gamma} \lambda \tau^2 + \tfrac{1}{6} e^{3 \gamma} \lambda^2 \tau^3+... \,$. The weak field limit $\lambda\to 0$ reduces to the CH function of the ModMax-like theory. By applying the CH function in ~\eqref{qbi}, we can derive the auxiliary field $\tau$ from identity ~\eqref{gensol} as shown below:\footnote{ProductLog$[z]$ gives the principal solution for $w$ in $z=w e^w$.}
\begin{eqnarray}\label{qbitau}
	\tau =  - \frac{e^{-\gamma}\,\, \text{ProductLog}\left[-2 \lambda \,\, e^{-\gamma + 2 e^{\gamma} \lambda q_2} \,\, q_1\right]}{2 \lambda} + q_2\,.
\end{eqnarray}
In the limit $\mathbf{q}\to \infty$ for the $\mathbf{q}$-deformed theory, the corresponding Lagrangian can be obtained as follows:
\begin{eqnarray}\label{Lqbitauq}
	\mathcal{L}_{\mathbf{q}\to \infty}& =  & \frac{1  }{\lambda}\Big[ 1 -  e^{\tfrac{1}{2} \text{ProductLog}\left[-2 \lambda  e^{-\gamma + 2 e^{\gamma} \lambda q_2} \,\, q_1\right] -  e^{\gamma} \lambda q_2}\\
	&& - 2   \lambda  q_1 e^{-\tfrac{1}{2} \text{ProductLog}\left[-2 \lambda  e^{-\gamma + 2 e^{\gamma} \lambda q_2}  q_1\right] - \gamma + e^{\gamma} \lambda q_2} \Big].\nonumber
\end{eqnarray}
We can write the expansion of the Lagrangian ~\eqref{Lqbitauq}  up to $\lambda^2$ as follows:
\begin{equation}\label{Expqbitau}
	\mathcal{L}_{\mathbf{q}\to \infty} =   - e^{-\gamma} q_1 + e^{\gamma} q_2 -  \tfrac{1}{2} e^{-2 \gamma} \lambda (q_1 + e^{2 \gamma} q_2)^2 + \tfrac{1}{6} e^{-3 \gamma} \lambda^2 (-5 q_1 + e^{2 \gamma} q_2) (q_1 + e^{2 \gamma} q_2)^2\,.
\end{equation}
Similar to other instances, the Lagrangian ~\eqref{Expqbitau} simplifies to the 2D  ModMax{-like}  Lagrangian ~\eqref{LMM} when approaching the limit of $\lambda \to 0$.

In general, we examined the q-deformed theories using the CH function ~\eqref{qDef} and solved equation ~\eqref{gensol} for the states $\mathbf{q}=\frac{3}{2}$, $\mathbf{q}= \frac{3}{4}$, and $\mathbf{q}\to \infty$. We derived the corresponding Lagrangians for these models and demonstrated that all these integrable models converge to the 2D ModMax{--like}  theory in the limit $\lambda \to 0$.
\subsection{Logarithmic integrable  model }\label{262}
In this section, we derive a new class of integrable theories from the CH approach, {  inspired by Logarithmic electromagnetism~\cite{Russo:2024csd} }, featuring a logarithmic structure distinct from the other theories discussed in this paper. To achieve this, we consider the CH function as follows:
\begin{eqnarray}\label{CHlog}
	\ell (\tau) = \frac{1}{\lambda} \log(1 + e^{\gamma} \lambda \tau)\,. 
\end{eqnarray}
The extension of the above logarithmic CH-function is as follows: 
\begin{eqnarray}\label{log}
	\ell (\tau)=e^{\gamma} \tau -  \tfrac{1}{2} e^{2 \gamma} \lambda \tau^2 + \tfrac{1}{3} e^{3 \gamma} \lambda^2 \tau^3+... \,.
\end{eqnarray}
Using the CH-function ~\eqref{CHlog}, we can find the auxiliary field $\tau$ for the two variables $q_1$ and $q_2$:
\begin{eqnarray}\label{loggg}
	\tau= \frac{1 - 2 e^{-\gamma} \lambda q_1 -  \sqrt{1 - 4 e^{-\gamma} \lambda q_1 - 4 \lambda^2 q_1 q_2}}{2 \lambda^2 q_1}\,.
\end{eqnarray}
The Lagrangian density is 
\begin{equation}\label{LLog}
	\mathcal{L}_{Log} =\frac{1}{\lambda}\bigg( \log\Bigl(\frac{e^{\gamma} }{2 \lambda q_1} \Bigl(1 -  \sqrt{1 -  4 \lambda e^{-\gamma} q_1 - 4 \lambda^2 q_1 q_2}\Bigr) \Bigr) + \sqrt{1 -  4 \lambda e^{-\gamma} q_1  - 4 \lambda^2 q_1 q_2} -1  \bigg)
\end{equation}

In this section, we employ the Courant-Hilbert approach to derive solutions of the differential equation ~\eqref{PDE2D}, which correspond to novel two-dimensional integrable theories. Our analysis yields several important examples, including not only the well-known ModMax and generalized Born-Infeld theories but also new q-deformed and logarithmic integrable systems. Building on these results, the subsequent section will establish a unified root flow equation that simultaneously describes both four-dimensional electrodynamic theories and their two-dimensional integrable counterparts. This universal formulation will then be rigorously verified against the specific solutions obtained in the current section, thereby demonstrating its consistency across dimensions and theoretical frameworks.

\section{Unique representation of root $T\bar{T}$   operator in $2D$ and $4D$}\label{05}
Generally, the energy-momentum tensor flow equations are both marginal and irrelevant in four-dimensional electrodynamic and two-dimensional scalar theories. It has been demonstrated that these flow equations are commutative in both two and four dimensions. We know that the solutions to PDE ~\eqref{dUdN4} in four dimensions are NEDs theories invariant under SO(2) transformations. 
For the two-dimensional case, we consider a theory involving $N \ge 1$ scalar fields $\Phi^i$ $(i=1,\dots, N)$, and the solutions to PDE~\eqref{dUdN} are integrable scalar theories invariant under $SO(N)\times SO(N)$ \cite{Babaei-Aghbolagh:2025lko}.
The PDEs ~\eqref{dUdN4} and ~\eqref{dUdN} represent a unified framework for two different groups of theories in four and two dimensions. In fact, equations ~\eqref{dUdN4} and ~\eqref{dUdN}  share the same nature. Given that these two equations share the same representation and their solutions are homogeneous using the CH solution approach, we anticipate that the flow equations of these differential equations will exhibit the same representation in two-dimensional and four-dimensional spaces. In this section, we derive the same representation for the root $T\bar{T}$ flow equation and the irrelevant $T\bar{T}$ flow equation in two and four dimensions.
\subsection{Energy-momentum tensor flow equations in duality invariant electromagnetic theories}\label{241}
To derive a new representation for the flow equations using the CH approach, we need to rewrite the energy-momentum tensor of the Lagrangian in ~\eqref{SPDE4} in terms of the CH functions $\ell$ and the variable $\tau$. The energy-momentum tensor (EMT) of Lagrangian ~\eqref{SPDE4}, which adheres to the duality condition from PDE ~\eqref{dUdN4}, has been derived as shown in \cite{Russo:2024xnh,Russo:2024dual}:
\begin{equation}\label{TMNlED4}
	T_{\mu\nu} = \left[\frac{\tau\dot\ell}{U+V}\right] T^{\rm Max}_{\mu\nu} +
	(\ell - \tau \dot\ell) {\rm g}_{\mu\nu} \, ,  
\end{equation}
where $T^{Max}_{\mu\nu}=F_{\mu\rho}{F_{\nu}}^{\rho}+ \,g_{\mu\nu} S$ is the (EMT) of the Maxwell theory.  Consequently, the trace of the  (EMT) in any self-dual electrodynamic theory is generally expressed as: 
\begin{equation}\label{TnNMM}
	{T_{\mu}}^\mu = 4(\ell - \tau \dot\ell)  \, .  
\end{equation}
The traceless condition for these theories is satisfied when $\ell = \tau \dot{\ell}$. The ModMax theory is the only nonlinear electrodynamic theory that meets this condition. The $T\bar{T}$ operators include the root $T\bar{T}$ operator $R_\gamma$ and the irrelevant $T\bar{T}$ operator $\mathcal{O}_{\lambda}$, which are special functions of two independent structures: ${T_{\mu}}^{\mu} {T_{\nu}}^{\nu}$ and $T_{\mu\nu}T^{\mu\nu}$. These structures can be expressed using Eq. ~\eqref{TMNlED4} as follows:
\begin{equation}\label{Tl2}   
	{T_{\mu}}^{\mu} {T_{\nu}}^{\nu}= 16(\ell - \tau\dot\ell)^2\, ,   
	\qquad
	T_{\mu\nu}T^{\mu\nu}  =  4\left[(\tau\dot\ell)^2 + (\ell -\tau\dot\ell)^2\right] \, .  
\end{equation}
As demonstrated in \cite{Babaei-Aghbolagh:2025cni}, using the two structures in ~\eqref{Tl2}, the irrelevant and marginal operators can be expressed in terms of the CH function as follows:
\begin{eqnarray}\label{Rto}
	&&\mathcal{R}_\gamma = \frac{1}{2}\, \sqrt{T_{\mu\nu}T^{\mu\nu}-\frac{1}{4} {T_{\mu}}^{\mu} {T_{\nu}}^{\nu}} =\tau\dot\ell (\tau)\,\,;\\
	&&\mathcal{O}_\lambda =\frac{1}{4}\, \Big(T_{\mu\nu}T^{\mu\nu}- \frac{1}{2} {T_{\mu}}^{\mu} {T_{\nu}}^{\nu}\Big) =- \ell (\tau)\,  \Big( \ell (\tau) -2 \tau\,\,\dot \ell (\tau) \Big)\,.\nonumber
\end{eqnarray}
Various electrodynamic theories, including Born-Infeld and logarithmic self-dual electromagnetic theories, are studied in detail in \cite{Babaei-Aghbolagh:2025cni}. Their irrelevant and marginal flow equations are investigated using the CH-function. 

This section examined the flow equations of four-dimensional self-dual electromagnetic theories using the CH approach. We demonstrated that this approach yields a new representation of these flow equations in ~\eqref{Rto}. In the next section, we extend this approach to two-dimensional integrable theories, showing that a universal representation exists for these flow equations, independent of the dimension of the theories.
\subsection{Energy-momentum tensor flow equations in integrable models}\label{221}
We can consider the standard definition for the energy-momentum tensor of Lagrangian $\mathcal{L}(q_1, q_2)$  as follows:
\begin{equation}
	\label{Tmunud}
	T_{\mu \nu}=\frac{-2}{ \sqrt{-g}}\frac{\partial\,   \sqrt{-g} \mathcal{L}(q_1 , q_2)  }{ \partial g^{\mu \nu}} = -2 \frac{\partial \mathcal{L}(q_1 , q_2)  }{ \partial g^{\mu \nu}} +g_{\mu \nu} \mathcal{L}(q_1 , q_2) \, . 
\end{equation}
Using chain  rule and the fact that $\frac{\partial \mathcal{L} }{ \partial q_1} = - \tfrac{1}{ \dot \ell (\tau)}$ and $\tfrac{\partial \mathcal{L}  }{ \partial q_2}=\dot \ell (\tau)$, we can obtain:
\begin{eqnarray}\label{CRule}
	\frac{\partial \mathcal{L}(q_1 , q_2)  }{ \partial g^{\mu \nu}}& =& \frac{\partial \mathcal{L}(q_1 , q_2)  }{ \partial q_1} \frac{\partial q_1 }{ \partial g^{\mu \nu}}+ \frac{\partial \mathcal{L}(q_1 , q_2)  }{ \partial q_2} \frac{\partial q_2  }{ \partial g^{\mu \nu}}=- \frac{1}{ \dot \ell (\tau)} \frac{\partial q_1 }{ \partial g^{\mu \nu}}+ \dot \ell (\tau) \frac{\partial q_2  }{ \partial g^{\mu \nu}}\\
	&=& - \frac{1}{ \dot \ell (\tau)} \biggl(\frac{\partial q_1 }{ \partial P_1}\frac{\partial P_1 }{ \partial g^{\mu \nu}}+\frac{\partial q_2  }{ \partial P_2}\frac{\partial P_2  }{ \partial g^{\mu \nu}}\biggl)+ \dot \ell (\tau) \biggl(\frac{\partial q_2  }{ \partial P_1}\frac{\partial P_1  }{ \partial g^{\mu \nu}}+\frac{\partial q_2  }{ \partial P_2}\frac{\partial P_2  }{ \partial g^{\mu \nu}}\biggl)\nonumber
\end{eqnarray}
We consider the energy-momentum tensor of the PCM as follows:
\begin{equation}
	\label{TPCM}
	{\tilde{\mathbf{T}}}_{\mu \nu}=\frac{-2}{ \sqrt{-g}}\frac{\partial\,   \sqrt{-g} \mathcal{L}_{PCM}  }{ \partial g^{\mu \nu}} = - \frac{\partial P_1  }{ \partial g^{\mu \nu}} +\frac{1}{ 2}  P_1 \, g_{\mu \nu} \, . 
\end{equation}
Using Eq. ~\eqref{TPCM}, we can derive the terms $\frac{\partial P_1}{\partial g^{\mu \nu}}$ and $\frac{\partial P_2}{\partial g^{\mu \nu}}$ as follows:
\begin{eqnarray}
	\label{TP1P2}
	\frac{\partial P_1  }{ \partial g^{\mu \nu}} &=&\frac{1}{ 2}  P_1 \,g_{\mu \nu}-{\tilde{\mathbf{T}}}_{\mu \nu}=- (q_1-q_2)\, g_{\mu \nu}-{\tilde{\mathbf{T}}}_{\mu \nu} \\
	\frac{\partial P_2  }{ \partial g^{\mu \nu}}&=& \frac{1}{ 2} \,P_1^2 \, g^{\mu \nu}  -2\, P_1 {\tilde{\mathbf{T}}}_{\mu \nu} +2\, {\tilde{\mathbf{T}}}_{\mu }{}^\alpha \,  {\tilde{\mathbf{T}}}_{\nu \alpha}\,\nonumber\\
	&=&2 \,(q_1-q_2)^2 \, g^{\mu \nu}  +4\, (q_1-q_2) \,{\tilde{\mathbf{T}}}_{\mu \nu} +2\, {\tilde{\mathbf{T}}}_{\mu }{}^\alpha \,  {\tilde{\mathbf{T}}}_{\nu \alpha}\nonumber \, . 
\end{eqnarray}
By substituting Eqs. ~\eqref{CRule} and ~\eqref{TP1P2} into Eq. ~\eqref{Tmunud} and using Eq. ~\eqref{gensol}, we can derive the energy-momentum tensor of the Lagrangian $\mathcal{L}(q_1 , q_2)$:
\begin{eqnarray}
	\label{Tmunu}
	T_{\mu \nu}&=&\biggl(\ell (\tau) -\tau \,   \dot \ell (\tau)  + \frac{q_1 + q_2}{2} \Big(  \dot \ell (\tau)- \frac{1}{ \dot \ell (\tau)} \Big) \biggr)  g_{\mu \nu} \\
	&& + \frac{\tau \, \dot \ell (\tau)  }{q_1 + q_2} {\tilde{\mathbf{T}}}_{\mu \nu} -  \frac{1 }{2 (q_1 + q_2)}\Big(   \dot \ell (\tau) - \frac{1}{ \dot \ell (\tau)}\Big) {\tilde{\mathbf{T}}}_{\mu}{}^{\alpha} {\tilde{\mathbf{T}}}_{\nu \alpha}\, . \nonumber
\end{eqnarray}
Using Eq. ~\eqref{TPCM}, it can be shown that:
\begin{equation}
	\label{TTTr}
	{\tilde{\mathbf{T}}_{\mu}}\,^{ \mu}	 =0\,\,, \,\,\,\,\,\,\,\,\,\,\,\, {\tilde{\mathbf{T}}}_{\mu\nu} {\tilde{\mathbf{T}}}^{\mu\nu} =2 (q_1+q_2)^2
\end{equation}
By applying Eqs. ~\eqref{TTTr} and ~\eqref{Tmunu},  we can obtain the two structures, $T_{\mu\nu}T^{\mu\nu}$ and ${T_{\mu}}^{ \mu}\,{T_{\nu}}^{ \nu}$, as follows:
\begin{equation}
	\label{Tnunu1}
	T_{\mu\nu}T^{\mu\nu}=2 \left[\Big(\tau\dot\ell(\tau)\Big)^2 + \Big(\ell(\tau) -\tau\dot\ell(\tau) \Big)^2\right]\,\,, \,\,\,\,\,\,\,\,\,\,\,\, {T_{\mu}}^{ \mu}\,{T_{\nu}}^{ \nu}=4 \left[\ell (\tau) -   \tau\,\,\dot \ell (\tau) \right]^2 
\end{equation}
We can derive the $T \bar{T}$ operators, including both the marginal and irrelevant operators, from the two structures $T_{\mu\nu}T^{\mu\nu}$ and ${T_{\mu}}^{ \mu}\,{T_{\nu}}^{ \nu}$ in Eq. ~\eqref{Tnunu1}  in terms of $\tau$ and the CH function $\ell (\tau)$. For these two operators, we have:
\begin{equation}\label{lM2}
	\mathcal{R}_\gamma =\frac{1}{\sqrt{2}}\, \sqrt{T_{\mu\nu}T^{\mu\nu}-\frac{1}{2} {T_{\mu}}^{\mu} {T_{\nu}}^{\nu}} = \tau\,\,\dot \ell (\tau)\, ,
\end{equation}
and
\begin{equation}\label{lM28}
	\mathcal{O}_\lambda =\frac{1}{2}\, \Big(T_{\mu\nu}T^{\mu\nu}- {T_{\mu}}^{\mu} {T_{\nu}}^{\nu}\Big) =- \ell (\tau)\,  \Big( \ell (\tau) -2 \tau\,\,\dot \ell (\tau) \Big)\,.
\end{equation}
This method for calculating the energy-momentum tensor in the CH approach is independent of dimension. Specifically, use Maxwell's energy-momentum tensor $T^{Max}_{\mu\nu}$ instead of the PCM energy-momentum tensor ${\tilde{\mathbf{T}}}_{\mu \nu}$, and replace the variables $q_1$ with $U$ and $q_2$ with $V$ in four dimensions. By applying this change of variables and using the equality $g^{\alpha\beta} T^{Max}_{\mu\alpha} T^{Max}_{\mu\beta} =  (U+V)^2 \,g_{\mu \nu}$, we arrive at exactly $T_{\mu \nu}$ in Eq. ~\eqref{TMNlED4}.
In the CH approach, the representation of the two operators $\mathcal{R}_\gamma$ and $\mathcal{O}_\lambda$ is identical in two dimensions for integrable theories and four dimensions for electromagnetic theories. Therefore, this approach allows us to find a universal representation for the operators $\mathcal{R}_\gamma$ and $\mathcal{O}_\lambda$.

\subsection{Universal representation of $T\bar{T}$ operators}\label{841}
By comparing two structures, $T_{\mu\nu}T^{\mu\nu}$ and ${T_{\mu}}^{ \mu}\,{T_{\nu}}^{ \nu}$, in two dimensions (as in Eq. ~\eqref{Tnunu1}) and four dimensions (as in Eq. ~\eqref{Tl2}), we can infer that this structure is unique for the $d$-dimensions. In $d$-dimensions, these two structures, $T_{\mu\nu}T^{\mu\nu}$ and ${T_{\mu}}^{ \mu}\,{T_{\nu}}^{ \nu}$, are as follows:
\begin{equation}
	\label{Tnunudi}
	T_{\mu\nu}T^{\mu\nu}=d \left[\Big(\tau\dot\ell(\tau)\Big)^2 + \Big(\ell(\tau) -\tau\dot\ell(\tau) \Big)^2\right]\,\,, \,\,\,\,\,\,\,\,\,\,\,\, {T_{\mu}}^{ \mu}\,{T_{\nu}}^{ \nu}=d^2 \left[\ell (\tau) -   \tau\,\,\dot \ell (\tau) \right]^2 
\end{equation}
We can represent the root operator 
$\mathcal{R}_\gamma$
and the irrelevant operator
$\mathcal{O}_\lambda$ in arbitrary dimensions as follows:
\begin{equation}\label{lM2di}
	\mathcal{R}_\gamma =\frac{1}{\sqrt{d}}\, \sqrt{T_{\mu\nu}T^{\mu\nu}-\frac{1}{d} {T_{\mu}}^{\mu} {T_{\nu}}^{\nu}}\,; \,\,\,\,\,\,\, \,\,\, \mathcal{O}_\lambda =\frac{1}{d}\, \Big(T_{\mu\nu}T^{\mu\nu}- \frac{2}{d}{T_{\mu}}^{\mu} {T_{\nu}}^{\nu}\Big)\,.
\end{equation}
By substituting the two structures, 
$T_{\mu\nu}T^{\mu\nu}$ and ${T_{\mu}}^{ \mu}\,{T_{\nu}}^{ \nu}$, from Eq. ~\eqref{Tnunudi} in  Eq. ~\eqref{lM2di}, we obtain a universal representation of the operators 
$\mathcal{R}_\gamma$  and $\mathcal{O}_\lambda$, which remain independent of dimension. This representation is given by:
\begin{equation}\label{lM28di}	\mathcal{R}_\gamma  = \tau\,\,\dot \ell (\tau)\,; \,\,\,\,\,\,\, \,\,\,  \,\,\, \mathcal{O}_\lambda =- \ell (\tau)\,  \Big( \ell (\tau) -2 \tau\,\,\dot \ell (\tau) \Big)\,.
\end{equation}
In the theories explored in this paper, the root flow equation applies to all cases with respect to $\gamma$ coupling, while the Born-Infeld-like theories follow the irrelevant flow equation. Consequently, in all theories, regardless of dimension, the root flow equation is as follows:
\begin{equation}\label{lM28dij}
\frac{\partial \mathcal{L}}{\partial \gamma}=\sqrt{\frac{1}{d}T_{\mu\nu}T^{\mu\nu}-\frac{1}{d^2} {T_{\mu}}^{\mu} {T_{\nu}}^{\nu}} = \tau\,\,\dot \ell (\tau)\,\,.
\end{equation}

\subsection{Explicit checks the universal root flow equation in integrable theories}\label{22ii}
In this section, we examine the intricacies of the flow equations within the integrable theories introduced in Section~\eqref{261}. These details confirm that the presented theories are true as expected for the root flow equation: $ \frac{\partial \mathcal{L}}{\partial \gamma}= \tau\,\dot \ell (\tau)$. To achieve this end, we begin with the general Born-Infeld-like theory and then delve into the other theories in detail.

\subsubsection{Root flow equation for general Born-Infeld{-like} theory}\label{2BI}
Building upon our established framework, this section systematically examines the Born-Infeld{-like} theory, a paradigmatic example of integrable systems previously derived in Section~\ref{30} through the CH approach. Following our development of a general flow equation for two-dimensional integrable theories in Section ~\eqref{221}, we now present a detailed derivation of the general Born-Infeld-like flow equations as a concrete validation of our methodology. This case study rigorously verifies the consistency of our comprehensive root-flow equation framework, introduced in Eq.~\eqref{lM28dij}, while also demonstrating its applicability to fundamental integrable models. Through this demonstration, we establish both the robustness of our general formulation and its precise correspondence with known integrable structures.
By substituting the CH-function from ~\eqref{lBI} into the general energy-momentum tensor framework ~\eqref{Tmunu}, we derive the energy-momentum tensor for general Born-Infeld{-like} theory as follows\footnote{Please note that the energy-momentum tensor ~\eqref{TBIN} can be derived directly from the general Born-Infeld{-like} Lagrangian ~\eqref{GMSP} using the $T_{\mu \nu}=\frac{-2}{ \sqrt{-g}}\frac{\partial\,   \sqrt{-g} \mathcal{L}_{GBI}(P_1 , P_2)  }{ \partial g^{\mu \nu}}$ definition.}:
\begin{eqnarray}
	\label{TBIN}
	T_{\mu \nu}&=&g_{\mu \nu} \frac{ -1 + \lambda \Bigl(\cosh(\gamma) (q_1 -  q_2) + (q_1 + q_2) \bigl(\lambda (q_1 + q_2) - 2 \sinh(\gamma)\bigr)\Bigr)}{\lambda \sqrt{1 - 2 \lambda e^{-\gamma} q_1} \sqrt{1 + 2 \lambda e^{\gamma} q_2}}- \frac{g_{\mu \nu} }{ \lambda } \\
    &&+ \frac{e^{-\gamma} (q_1 + e^{2 \gamma} q_2) {\tilde{\mathbf{T}}}_{\mu \nu}}{\sqrt{1 -  2 \lambda e^{-\gamma} q_1} (q_1 + q_2) \sqrt{1 + 2 \lambda e^{\gamma} q_2}} + \frac{\bigl(\lambda (q_1 + q_2) -  \sinh(\gamma)\bigr) {\tilde{\mathbf{T}}}_{\mu}^{\alpha } {\tilde{\mathbf{T}}}_{\nu \alpha}}{\sqrt{1 -  2 \lambda e^{-\gamma} q_1} (q_1 + q_2) \sqrt{1 + 2 \lambda e^{\gamma} q_2}}\nonumber
\end{eqnarray}
We can explicitly calculate the structure $T_{\mu\nu}T^{\mu\nu}$ from the energy-momentum tensor of the general Born-Infeld-like theory in ~\eqref{TBIN}. Using ~\eqref{TBIN} and ~\eqref{TTTr}, we find for this structure:
 \begin{equation}
	\label{TTBIT}
	T_{\mu\nu}T^{\mu\nu}=\frac{4  (1 -  \lambda e^{-\gamma} q_1 + e^{ \gamma} \lambda q_2) \bigl(1-  \lambda e^{-\gamma}q_1 + e^{ \gamma} \lambda q_2 -   \sqrt{1 - 2 e^{-\gamma} \lambda q_1} \sqrt{1 + 2 e^{\gamma} \lambda q_2}\bigr)}{\lambda^2 (1 - 2 e^{-\gamma} \lambda q_1) (1 + 2 e^{\gamma} \lambda q_2)}
\end{equation}  
We can verify that by substituting the CH-function ~\eqref{lBI} into the general equation we derived for the $T_{\mu\nu}T^{\mu\nu}$
structure in the first equation of  ~\eqref{Tnunu1}, we precisely obtain Eq. ~\eqref{TTBIT}. We can also derive the following double-trace structure  from the energy-momentum tensor of the  general Born-Infeld-like theory in ~\eqref{TBIN} as follows:
\begin{equation}
	\label{TRBIT}
	 {T_{\mu}}^{ \mu}\,{T_{\nu}}^{ \nu}=\frac{4 \bigl(1 -  e^{-\gamma} \lambda q_1 + e^{\gamma} \lambda q_2 -  \sqrt{1 - 2 e^{-\gamma} \lambda q_1} \sqrt{1 + 2 e^{\gamma} \lambda q_2}\bigr)^2}{\lambda^2 (1 - 2 e^{-\gamma} \lambda q_1) (1 + 2 e^{\gamma} \lambda q_2)}
\end{equation}
To explicitly verify the root flow equation for Born-Infeld{-like} theory, we must derive the left-hand side of Equation ~\eqref{lM28dij} from Lagrangian ~\eqref{GMSP}, which yields:
\begin{equation}
	\label{DLBITg}
	\frac{\partial \mathcal{L}_{GBI}}{\partial \gamma}=\frac{e^{-\gamma} q_1 + e^{\gamma} q_2}{\sqrt{1 - 2 e^{-\gamma} \lambda q_1} \sqrt{1 + 2 e^{\gamma} \lambda q_2}}
\end{equation}
It can be readily demonstrated that the aforementioned flow equation is satisfied by equality $\frac{\partial \mathcal{L}_{GBI}}{\partial \gamma}=\tau \dot \ell\,$. Alternatively, we can derive the right-hand side of Equation ~\eqref{lM28dij} using structures ~\eqref{TTBIT} and ~\eqref{TRBIT}, yielding the following expression:
\begin{equation}
	\label{DLBITgp1}
	\frac{1}{\sqrt{2}}\, \sqrt{T_{\mu\nu}T^{\mu\nu}-\frac{1}{2} {T_{\mu}}^{\mu} {T_{\nu}}^{\nu}}=\frac{e^{-\gamma} q_1 + e^{\gamma} q_2}{\sqrt{1 - 2 e^{-\gamma} \lambda q_1} \sqrt{1 + 2 e^{\gamma} \lambda q_2}}\,.
\end{equation}
Finally, by comparing expressions  ~\eqref{DLBITg} and ~\eqref{DLBITgp1}, we can clearly demonstrate that the root flow equation  ~\eqref{lM28dij} holds for the two-dimensional generalized Born-Infeld{-like} integrable theory:
\begin{equation}
	\label{DLBITgp}
	\frac{\partial \mathcal{L}_{GBI}}{\partial \gamma}=\frac{1}{\sqrt{2}}\, \sqrt{T_{\mu\nu}T^{\mu\nu}-\frac{1}{2} {T_{\mu}}^{\mu} {T_{\nu}}^{\nu}}= \tau \dot \ell\,.
\end{equation}
In this section, we have rigorously verified the unique root flow equation ~\eqref{lM28dij} for two-dimensional Born-Infeld-like theory. Subsequently, we will analyze additional examples, including one of the $\mathbf{q}$-deform theories derived in Section ~\eqref{25a} and the logarithmic integrable theory presented in Section ~\eqref{262}, by Lagrangian density ~\eqref{LLog}.
\subsubsection{Root flow equation for $\mathbf{q}$-deformed theories}\label{2qdef1}
In Section ~\eqref{25a}, we conducted a detailed investigation of integrable q-deformed theories. The Lagrangians derived from these theories incorporate three distinct states: $\mathbf{q}=\frac{3}{2}$, $\mathbf{q}=\frac{3}{4}$, and $\mathbf{q} \to \infty$, as respectively established in Lagrangians ~\eqref{NoMq}, ~\eqref{qdef}, and ~\eqref{Lqbitauq}. Through explicit derivation, we demonstrate that the root flow equation for the state $\mathbf{q}=\frac{3}{2}$ satisfies precisely the unique form presented in ~\eqref{lM28dij}. For this purpose, we derive the energy-momentum tensor of $\mathbf{q}=\frac{3}{2}$-deformed theory  by substituting the CH-function of ~\eqref{q32} into the general form ~\eqref{Tmunu}, expressed as a function of  $q_1$ and $q_2$:
\begin{eqnarray}
	\label{TqAN}
	T_{\mu \nu}&=& g_{\mu \nu} \Bigl(\frac{1}{\lambda} -  \frac{e^{-\frac{ 3 \gamma}{4}} (5 q_1 + 3 q_2)}{\sqrt{6} \Gamma^{1/2}} + \frac{e^{-\frac{ \gamma}{4}} \Gamma^{1/2} \bigl(-6 + e^{\gamma} \lambda (3 q_1 + q_2)\bigr)}{6 \sqrt{6} \lambda}\Bigr) \\
    &+&  \bigl(\frac{\sqrt{6} e^{-3/4 \gamma} q_1}{\Gamma^{1/2} (q_1 + q_2)} + \frac{e^{\frac{ 3 \gamma}{4}} \Gamma^{1/2} q_2}{\sqrt{6} (q_1 + q_2)}\bigr) {\tilde{\mathbf{T}}}_{\mu \nu} + \bigl(\frac{\sqrt{3} e^{-\frac{ 3 \gamma}{4}}}{\sqrt{2} \Gamma^{1/2} (q_1 + q_2)} -  \frac{e^{\frac{ 3 \gamma}{4}} \Gamma^{1/2}}{2 \sqrt{6} (q_1 + q_2)}\bigr) {\tilde{\mathbf{T}}}_{\mu}{}^{\alpha } {\tilde{\mathbf{T}}}_{\nu \alpha} \,. \nonumber
\end{eqnarray}
Using the energy-momentum tensor of the $\mathbf{q}=\frac{3}{2}$-deformed theory from Eq. ~\eqref{TqAN}, we explicitly calculate the two structures, 
$T_{\mu\nu}T^{\mu\nu}$ and  ${T_{\mu}}^{ \mu}\,{T_{\nu}}^{ \nu}$, as follows:
 \begin{eqnarray}
	\label{TTqAT}
	T_{\mu\nu}T^{\mu\nu}&=&\frac{2}{\lambda^2} -  \frac{\sqrt{6} e^{-\frac{ \gamma}{4}} \Gamma^{1/2}}{\lambda^2} + \frac{e^{-\frac{3 \gamma}{4}} \Gamma^{3/2}}{3 \sqrt{6} \lambda^2} + q_1 (\frac{14 e^{-\gamma}}{3 \lambda} + \frac{20}{9} q_2)\\
    &+&\Gamma \, \Big(\frac{e^{-\frac{ \gamma}{2}}}{3 \lambda^2} -  \frac{5 e^{-\frac{ 3 \gamma}{2}} q_1}{9 \lambda} + \frac{2 e^{1/2 \gamma} q_2}{9 \lambda} + \tfrac{10}{27} e^{3/2 \gamma} q_2^2 \Big) \,, \nonumber
\end{eqnarray}  
and
\begin{equation}
	\label{TRqAT}
	 {T_{\mu}}^{ \mu}\,{T_{\nu}}^{ \nu}=\frac{\bigl(72 + \sqrt{6} e^{-\frac{3 \gamma}{4}} \Gamma^{1/2} (-18 e^{\frac{ \gamma}{2}} + \Gamma)\bigr)^2}{1296 \lambda^2}\,.
\end{equation}
By taking the derivative of Lagrangian ~\eqref{NoMq} with respect to $\gamma$, we derive the following expression:
\begin{equation}
	\label{DLqATg}
	\frac{\partial \mathcal{L}_{\mathbf{q}=\frac{3 }{2}}}{\partial \gamma}=\frac{\sqrt{3} e^{-\frac{ \gamma}{4}} \Gamma^{1/2}}{2 \sqrt{2} \lambda} -  \frac{e^{-\frac{3 \, \gamma}{4} } \Gamma^{3/2}}{4 \sqrt{6} \lambda}\,.
\end{equation}
Using the two structures ~\eqref{TTqAT} and ~\eqref{TRqAT}, we construct the root operator ~\eqref{lM2}. By comparing this operator with the Lagrangian derivative with respect to $\gamma$ in ~\eqref{DLqATg}, we demonstrate: 
\begin{equation}
	\label{DLqAa}
	\frac{\partial \mathcal{L}_{\mathbf{q}=\frac{3 }{2}}}{\partial \gamma}=\frac{1}{\sqrt{2}}\, \sqrt{T_{\mu\nu}T^{\mu\nu}-\frac{1}{2} {T_{\mu}}^{\mu} {T_{\nu}}^{\nu}}= \tau \dot \ell\,.
\end{equation}
The flow equation ~\eqref{DLqAa} shows that the structure of the $\mathbf{q}$-deformed theory by $\mathbf{q}=\frac{3 }{2}$, is fully consistent with the unique representation of the root flow equation ~\eqref{lM28dij}. Through direct computational verification, we have confirmed that additional $\mathbf{q}$-deformed theories, as described by Lagrangians ~\eqref{qdef}, and ~\eqref{Lqbitauq}, similarly satisfy the universal representation of the root flow equation.
\subsubsection{Root flow equation for Logarithmic theory}\label{2logt}
Building on the logarithmic integrable theory developed in Section ~\eqref{262} (see Eq. ~\eqref{LLog}), we now present a detailed derivation of its root flow equation, explicitly demonstrating its consistency with the general flow structure given in Eq. ~\eqref{lM28dij}. Our analysis begins with the Lagrangian formulation Eq. ~\eqref{LLog}, from which we systematically derive the corresponding energy-momentum tensor as follows:
\begin{eqnarray}
	\label{TqLoAN}
	T_{\mu \nu}&=& g_{\mu \nu} \Bigg(\frac{1}{\lambda} \Big(\log\bigl(\frac{e^{\gamma} (1 -  \sqrt{1 - 4 e^{-\gamma} \lambda q_1 - 4 \lambda^2 q_1 q_2})}{2 \lambda q_1}\bigr) -  \frac{3 q_1 + q_2}{2 q_1} \Big) \\
    &+& \frac{\lambda (q_1 + q_2)^2 + e^{-\gamma} (3 q_1 + q_2)}{1 -  \sqrt{1 - 4 e^{-\gamma} \lambda q_1 - 4 \lambda^2 q_1 q_2}}\Bigg) + \frac{1 - \sqrt{1 - 4 e^{-\gamma} \lambda q_1 - 4 \lambda^2 q_1 q_2} + 2 e^{\gamma} \lambda q_2 }{2 \lambda (q_1 + q_2) (1 + e^{\gamma} \lambda q_2)}  {\tilde{\mathbf{T}}}_{\mu \nu}\nonumber \\
    &+& \Big(\frac{1}{2 \lambda q_1^2 + 2 \lambda q_1 q_2} -  \frac{\lambda + \frac{e^{-\gamma}}{q_1 + q_2}}{1 -  \sqrt{1 - 4 e^{-\gamma} \lambda q_1 - 4 \lambda^2 q_1 q_2}}\Big) {\tilde{\mathbf{T}}}_{\mu}{}^{\alpha } {\tilde{\mathbf{T}}}_{\nu \alpha}\nonumber
\end{eqnarray}
Consistent with the integrable models analyzed throughout this work, we derive two fundamental structures from the energy-momentum tensor:
 \begin{eqnarray}
	\label{TTLoAT}
	T_{\mu\nu}T^{\mu\nu}&=&\frac{1}{\lambda^2} \Big( \frac{4 e^{-2 \gamma} \bigl(e^{\gamma} (1 -  \sqrt{1 - 4 e^{-\gamma} \lambda q_1 - 4 \lambda^2 q_1 q_2}) - 2 \lambda q_1\bigr)^2}{(1 -  \sqrt{1 - 4 e^{-\gamma} \lambda q_1 - 4 \lambda^2 q_1 q_2})^2} \\
    &+& 2 \Bigl(\log\bigl(\frac{e^{\gamma} (1 -  \sqrt{1 - 4 e^{-\gamma} \lambda q_1 - 4 \lambda^2 q_1 q_2})}{2 \lambda q_1}\bigr)\Bigr)^2  \nonumber \\
    &+&\log\bigl(\frac{e^{\gamma} (1 -  \sqrt{1 - 4 e^{-\gamma} \lambda q_1 - 4 \lambda^2 q_1 q_2})}{2 \lambda q_1}\bigr) ( \frac{8 e^{-\gamma} \lambda q_1}{1 -  \sqrt{1 - 4 e^{-\gamma} \lambda q_1 - 4 \lambda^2 q_1 q_2}}-4) \Big)\nonumber \, ,
\end{eqnarray}  
and 
\begin{equation}
	\label{TRLoAT}
	 {T_{\mu}}^{ \mu}\,{T_{\nu}}^{ \nu}=\Bigl( \frac{2 }{\lambda} \log\bigl(\frac{ 1 - \sqrt{1 - 4 e^{-\gamma} \lambda q_1 - 4 \lambda^2 q_1 q_2}}{2 \lambda e^{-\gamma} q_1}\bigr) - \frac{2}{\lambda} + \frac{4 e^{-\gamma} q_1}{1 -  \sqrt{1 - 4 e^{-\gamma} \lambda q_1 - 4 \lambda^2 q_1 q_2}}\Bigr)^2\,.
\end{equation}
Following the same procedure, we derive the following result by differentiating the logarithmic Lagrangian with respect to the coupling $\gamma$:
\begin{equation}
	\label{DLLoATg}
	\frac{\partial \mathcal{L}_{Log}}{\partial \gamma}= \frac{1 + 2 e^{\gamma} \lambda q_2 -  \sqrt{1 - 4 e^{-\gamma} \lambda q_1 - 4 \lambda^2 q_1 q_2}}{2 \lambda (1 + e^{\gamma} \lambda q_2)}\,.
\end{equation}
To conclude our analysis of the logarithmic theory, we rigorously verify that the root flow equation  ~\eqref{lM28dij} is satisfied, thereby demonstrating the consistency of our framework. This explicit validation confirms that the proposed theory maintains integrability under the specified root flow equation.
\begin{equation}
	\label{DLLoAa}
	\frac{\partial \mathcal{L}_{Log}}{\partial \gamma}=\frac{1}{\sqrt{2}}\, \sqrt{T_{\mu\nu}T^{\mu\nu}-\frac{1}{2} {T_{\mu}}^{\mu} {T_{\nu}}^{\nu}}\,.
\end{equation}
\section{Generalized solution of  integrable  sigma models}\label{ff8412ff2}
The root flow equation originates from the free theory where the uncoupled parameter $\lambda$ vanishes, while its interacting counterpart emerges from the $\lambda$-dependent terms. Consequently, the choice of an appropriate free theory, whose deformation via interaction terms yields the desired structure, is essential.
In this section, we focus on deformed theories that reduce to the two-dimensional integrable ModMax{-like}  theory in the non-interacting limit. To generate such integrable theories systematically, we consider the associated CH-function, expanded as follows:
\begin{equation}\label{Lexpan}
	\ell (\tau) =e^\gamma \,  \tau  +\sum_{i=1}^\infty  \, \lambda^if_i(\gamma)  \, \tau^{i+1}\,,
\end{equation}
where the $f_i(\gamma)$ are functions of $\gamma$. In practice, this choice encompasses a broad class of integrable theories. However, since obtaining exact solutions to the second equation in~\eqref{gensol} and determining the corresponding $\tau$-function at arbitrary order in $\lambda$ proves intractable, we instead focus on perturbative calculations up to $\mathcal{O}(\lambda^3)$. The general theory expansion takes the following form:
\begin{equation}\label{Lexpan}
	\ell (\tau) =e^\gamma \,  \tau  +\lambda \,  f_1(\gamma) \,  \tau^{2}  +\lambda^2 \,  f_2(\gamma) \, \tau^{3}  +\lambda^3 \, f_3(\gamma) \,  \tau^{4} +...\,\,.
\end{equation}
To derive a flow equation consistent with the ModMax-like theory in the deformed theory, we need to consider how the functions $f_i(\gamma)$ depend on the deformation parameter $\gamma$. We begin by determining the Lagrangian corresponding to the CH-function up to the third order in $\lambda^3$. Subsequently, by imposing constraints derived from the flow equation, we obtain the $\gamma$ dependence of the coefficient functions. By employing the CH-function~\eqref{Lexpan}, we solve the second equation in Eq.~\eqref{gensol} and derive the $\tau$-form up to order in $\lambda^3$, yielding:
\begin{eqnarray}\label{Gqqwn11}
	\tau&=& e^{-2 \gamma} q_1 + q_2 - 4\, \lambda \, e^{-5 \gamma} \,f_1(\gamma) \,q_1\, (q_1 + e^{2 \gamma} q_2)\\
    &+& 2\,\lambda^2\,  e^{-8 \gamma} \, q_1 (q_1 + e^{2 \gamma} q_2) \,\Bigl(-3 e^{\gamma} f_2(\gamma) (q_1 + e^{2 \gamma} q_2) + 2 \bigl(f_1(\gamma)\bigr)^2 (7 q_1 + 3 e^{2 \gamma} q_2)\Bigr)\nonumber \\
    &-& 4\,\lambda^3\, e^{-11 \gamma} \, q_1 (q_1 + e^{2 \gamma} q_2) \,\Bigl(2 e^{2 \gamma} f_3(\gamma) (q_1 + e^{2 \gamma} q_2)^2 \nonumber \\
    &-&9 e^{\gamma} f_1(\gamma) f_2(\gamma) (q_1 + e^{2 \gamma} q_2) (3 q_1 + e^{2 \gamma} q_2)+ 4 \bigl(f_1(\gamma)\bigr)^3 (5 q_1 + e^{2 \gamma} q_2) (3 q_1 + 2 e^{2 \gamma} q_2)\Bigr) \nonumber\,.
\end{eqnarray}
Using~\eqref{Lexpan} and ~\eqref{Gqqwn11}, we can obtain ($\dot{\ell}(\tau)$) and then the deformed Lagrangian to order $\lambda^3$, which corresponds to the CH-function~\eqref{Lexpan}, is obtained as follows:
\begin{eqnarray}\label{Gqqn11q}
	\mathcal{L}_{GSM} (q_1,q_2)&=& - e^{-\gamma} q_1 + e^{\gamma} q_2 + e^{-4 \gamma} \lambda f_1(\gamma) (q_1 + e^{2 \gamma} q_2)^2 \nonumber \\
    &+& e^{-7 \gamma} \lambda^2 (q_1 + e^{2 \gamma} q_2)^2 \Bigl(-4 \bigl(f_1(\gamma)\bigr)^2 q_1 + e^{\gamma} f_2(\gamma) (q_1 + e^{2 \gamma} q_2)\Bigr) \nonumber \\
    &+& e^{-10 \gamma} \lambda^3 (q_1 + e^{2 \gamma} q_2)^2 \Bigl(-12 e^{\gamma} f_1(\gamma) f_2(\gamma) q_1 (q_1 + e^{2 \gamma} q_2) \nonumber \\
    &+& e^{2 \gamma} f_3(\gamma) (q_1 + e^{2 \gamma} q_2)^2 + 8 \bigl(f_1(\gamma)\bigr)^3 q_1 (3 q_1 + e^{2 \gamma} q_2)\Bigr)\,.
\end{eqnarray}
The above Lagrangian encompasses a broad class of theories featuring two coupling constants, $\gamma$ and $\lambda$, derived from the CH-functions in Eq.~\eqref{Lexpan}.  Using Eq.~\eqref{UN}, we derive the Lagrangian~\eqref{Gqqn11q} expressed in terms of two fundamental variables $P_1$ and $P_2$ as follows:
\begin{eqnarray}\label{Gqqdn11pp}
	\mathcal{L}_{GSM}&=& + \tfrac{1}{2} \bigl(\cosh(\gamma) P_1 + \sqrt{2P_2-P_1^2} \sinh(\gamma)\bigr) \\
    &+&\tfrac{1}{4} e^{-2 \gamma} \lambda f_1(\gamma) \bigl(\cosh(\gamma) \sqrt{2P_2-P_1^2} + P_1 \sinh(\gamma)\bigr)^2  \nonumber \\
    &+&\lambda^2 \Bigl(- \tfrac{1}{4} e^{-5 \gamma} \bigl(f_1(\gamma)\bigr)^2 \bigl(- P_1 + \sqrt{2P_2-P_1^2}\bigr) \bigl(\cosh(\gamma) \sqrt{2P_2-P_1^2} + P_1 \sinh(\gamma)\bigr)^2  \nonumber \\
    &+& \tfrac{1}{8} e^{-3 \gamma} f_2(\gamma) \bigl(\cosh(\gamma) \sqrt{2P_2-P_1^2} + P_1 \sinh(\gamma)\bigr)^3\Bigr)  \nonumber \\
    &+&\lambda^3 \Bigl(\tfrac{1}{8} e^{-8 \gamma} \bigl(f_1(\gamma)\bigr)^3 \bigl(- P_1 + \sqrt{2P_2-P_1^2}\bigr) \bigl((-3 + e^{2 \gamma}) P_1 + (3 + e^{2 \gamma}) \sqrt{2P_2-P_1^2}\bigr) \nonumber \\
    &\times& \bigl(\cosh(\gamma) \sqrt{2P_2-P_1^2} + P_1 \sinh(\gamma)\bigr)^2 -  \tfrac{3}{8} e^{-6 \gamma} f_1(\gamma) f_2(\gamma) \bigl(- P_1 + \sqrt{2P_2-P_1^2}\bigr)  \nonumber \\
    &\times&\bigl(\cosh(\gamma) \sqrt{2P_2-P_1^2} + P_1 \sinh(\gamma)\bigr)^3 + \tfrac{1}{16} e^{-4 \gamma} f_3(\gamma) \bigl(\cosh(\gamma) \sqrt{2P_2-P_1^2} + P_1 \sinh(\gamma)\bigr)^4\Bigr)\nonumber\,.
\end{eqnarray}
By appropriately specifying the $\gamma$-dependence of the functions $f_i(\gamma)$, the Lagrangian in Eq.~\eqref{Gqqn11q} can be made compatible with the flow equation with respect to the coupling $\gamma$.
To achieve this, we proceed by computing the derivative of the Lagrangian~\eqref{Gqqn11q} with respect to $\gamma$. Differentiating, we obtain:
\begin{eqnarray}\label{Gqqjjn11}
	\frac{\partial \mathcal{L}_{GSM}}{\partial \gamma}&=&e^{-\gamma} q_1 + e^{\gamma} q_2 + \lambda \bigl(-4 e^{-4 \gamma} f_1(\gamma) q_1 (q_1 + e^{2 \gamma} q_2) + e^{-4 \gamma} f^{\prime}_1(\gamma) (q_1 + e^{2 \gamma} q_2)^2\bigr)   \\
    &+& \lambda^2 \Bigl(-8 e^{-7 \gamma} f^{\prime}_1(\gamma) f_1(\gamma) q_1 (q_1 + e^{2 \gamma} q_2)^2 - 6 e^{-6 \gamma} f_2(\gamma) q_1 (q_1 + e^{2 \gamma} q_2)^2  \nonumber \\
    &+&  e^{-6 \gamma} f^{\prime}_2(\gamma) (q_1 + e^{2 \gamma} q_2)^3 +4 e^{-7 \gamma} \bigl(f_1(\gamma)\bigr)^2 q_1 (q_1 + e^{2 \gamma} q_2) (7 q_1 + 3 e^{2 \gamma} q_2)\Bigr)  \nonumber \\
    &+& \lambda^3 \Bigl(-12 e^{-9 \gamma} f^{\prime}_1(\gamma) f_2(\gamma) q_1 (q_1 + e^{2 \gamma} q_2)^3 - 8 e^{-8 \gamma} f_3(\gamma) q_1 (q_1 + e^{2 \gamma} q_2)^3  \nonumber \\
    &+&  e^{-8 \gamma} f^{\prime}_3(\gamma) (q_1 + e^{2 \gamma} q_2)^4 + 24 e^{-10 \gamma} f^{\prime}_1(\gamma) \bigl(f_1(\gamma)\bigr)^2 q_1 (q_1 + e^{2 \gamma} q_2)^2 (3 q_1 + e^{2 \gamma} q_2)  \nonumber \\
    &-&  16 e^{-10 \gamma} \bigl(f_1(\gamma)\bigr)^3 q_1 (q_1 + e^{2 \gamma} q_2) (5 q_1 + e^{2 \gamma} q_2) (3 q_1 + 2 e^{2 \gamma} q_2)  \nonumber \\
    &+&  f_1(\gamma) \bigl(-12 e^{-9 \gamma} f^{\prime}_2(\gamma) q_1 (q_1 + e^{2 \gamma} q_2)^3 + 36 e^{-9 \gamma} f_2(\gamma) q_1 (q_1 + e^{2 \gamma} q_2)^2 (3 q_1 + e^{2 \gamma} q_2)\bigr)\Bigr)\nonumber
\end{eqnarray}
In the remainder of this section, we determine the appropriate $\gamma$-dependence for the functions $f_i$ in the Lagrangian~\eqref{Gqqn11q} to derive perturbatively integrable theories compatible with the root flow equations~\eqref{lM28dij}. Furthermore, we demonstrate that these integrable theories can incorporate flow equations proportional to the CH-function. We show that these new flow equations constitute a rescaling of the root flow equations.
\subsection{Root flow equation for general perturbation theory }\label{021}
In this section, we study the flow equations of the Lagrangian~\eqref{Gqqn11q} with respect to $\gamma$ using the general form of the $T\bar{T}$-root deformation. We determine the functions $f_i(\gamma)$ in terms of $\gamma$ and demonstrate that, in a special case, this general root flow equation reduces to the standard root flow equation in two dimensions. In this particular case, the flow equation~\eqref{lM28dij} holds for the deformed integrable theory. 
We consider a two-dimensional root flow equation in the following general form:
\begin{equation}\label{ldijnn}
    \frac{\partial \mathcal{L}}{\partial \gamma} =  \sqrt{ e_1 \, T_{\mu\nu} T^{\mu\nu} - e_2 \, {T_{\mu}}^{\mu} {T_{\nu}}^{\nu} } \,.
\end{equation}
where and $T_{\mu\nu}$ represents the energy-momentum tensor of the general deformed theory in \eqref{Gqqn11q}. The two structures $T_{\mu\nu} T^{\mu\nu}$ and ${T_{\mu}}^{\mu} {T_{\nu}}^{\nu}$ can be derived from the energy-momentum tensor of the deformed theory~\eqref{Gqqn11q} as follows:
\begin{eqnarray}\label{Gqqn11}
T_{\mu\nu} T^{\mu\nu} &=& 2 e^{-2 \gamma} (q_1 + e^{2 \gamma} q_2)^2 + 8 e^{-5 \gamma} \lambda f_1(\gamma) (- q_1 + e^{2 \gamma} q_2) (q_1 + e^{2 \gamma} q_2)^2 \\
    &+& \lambda^2 \Bigl(12 e^{-7 \gamma} f_2(\gamma) (- q_1 + e^{2 \gamma} q_2) (q_1 + e^{2 \gamma} q_2)^3 \nonumber\, \\
    &+& 2 e^{-8 \gamma} \bigl(f_1(\gamma)\bigr)^2 (q_1 + e^{2 \gamma} q_2)^2 (29 q_1^2 - 14 e^{2 \gamma} q_1 q_2 + 5 e^{4 \gamma} q_2^2)\Bigr) \nonumber\,\\
    &+& \lambda^3 \Bigl(-512 e^{-11 \gamma} \bigl(f_1(\gamma)\bigr)^3 q_1^3 (q_1 + e^{2 \gamma} q_2)^2 + 16 e^{-9 \gamma} f_3(\gamma) (- q_1 + e^{2 \gamma} q_2) (q_1 + e^{2 \gamma} q_2)^4 \nonumber\, \\
    &+& 32 e^{-10 \gamma} f_1(\gamma) f_2(\gamma) (q_1 + e^{2 \gamma} q_2)^3 (7 q_1^2 - 4 e^{2 \gamma} q_1 q_2 + e^{4 \gamma} q_2^2)\Bigr)  \nonumber\,.
\end{eqnarray}
and
\begin{eqnarray}\label{Gqqn1tr1}
 {T_{\mu}}^{\mu} {T_{\nu}}^{\nu} &=& 4 e^{-8 \gamma} \lambda^2 \bigl(f_1(\gamma)\bigr)^2 (q_1 + e^{2 \gamma} q_2)^4  \\
    &+& \lambda^3 \Bigl(-64 e^{-11 \gamma} \bigl(f_1(\gamma)\bigr)^3 q_1 (q_1 + e^{2 \gamma} q_2)^4 + 16 e^{-10 \gamma} f_1(\gamma) f_2(\gamma) (q_1 + e^{2 \gamma} q_2)^5\Bigr) \nonumber\,.
\end{eqnarray}
Using these two structures, we can construct the right-hand side of the flow equation~\eqref{ldijnn}, while the left-hand side appears in equation~\eqref{Gqqjjn11}. The right-hand side of the flow equation involves the functions $f_i(\gamma)$, whereas the left-hand side contains their derivatives $f_i'(\gamma)$.  By comparing the flow equation~\eqref{ldijnn} order by order, we can formulate a differential equation to determine the $f_i(\gamma)$ functions and express them explicitly in terms of $\gamma$.
The root flow equation originates from the zeroth order in the uncoupled parameter $\lambda$ and does not involve any $f_i(\gamma)$ functions. At this order, only a constant coefficient $e_1$ appears, which is determined to be $e_1 = \frac{1}{2}$ by matching both sides of the flow equation~\eqref{ldijnn}. 
Applying this constant value $e_1 = \frac{1}{2}$, we compare the sides of the general flow equation~\eqref{ldijnn} at order $\lambda$, yielding the following differential equation:
\begin{eqnarray}\label{df01}
 f^{\prime}_1(\gamma)- 2 f_1(\gamma)=0\,.
\end{eqnarray}
By solving the differential equation above, we determine the function $f_1(\gamma)$ to be:
\begin{eqnarray}\label{df011}
 f_1(\gamma)=m_1  e^{2 \gamma}\,.
\end{eqnarray}
where $m_1$ is a constant.
By substituting $f_1(\gamma)$, we can compare both sides of the flow equation~\eqref{ldijnn} in order $\mathcal{O}(\lambda^2)$. This comparison yields the following differential equation for $f_2(\gamma)$:
\begin{eqnarray}\label{df02}
 f^{\prime}_2(\gamma) - \frac{1}{2} e^{3 \gamma} (1- 4 e_2) m_1^2 - 3 f_2(\gamma)=0\,.
\end{eqnarray}
Solving this differential equation yields the function $f_2(\gamma)$:
\begin{eqnarray}\label{df022}
 f_2(\gamma)= m_2 e^{3 \gamma}  + \frac{1}{2} e^{3 \gamma} (1 - 4 e_2)\, m_1^2 \,  \gamma\,,
\end{eqnarray}
where $m_2$ is a constant. By iterating this procedure, we derive the following differential equation for $f_3(\gamma)$ at $\mathcal{O}(\lambda^3)$:
\begin{eqnarray}\label{df03}
 f^{\prime}_3(\gamma) + e^{4 \gamma} (1 - 4 e_2) m_1 \bigl(-2 m_2 + m_1^2 (1 -  \gamma + 4 e_2 \gamma)\bigr) - 4 f_3(\gamma)=0\,.
\end{eqnarray}
Solving the differential equation for $f_3(\gamma)$ yields the solution:
\begin{eqnarray}\label{df033}
 f_3(\gamma)=m_3\, e^{4 \gamma}  - e^{4 \gamma} (1 - 4 e_2) m_1 \bigl( (m_1^2  - 2 m_2 )\gamma -\tfrac{1}{2} (1 - 4 e_2) m_1^2 \gamma^2\bigr)\,.
\end{eqnarray}
Therefore, we have systematically determined the functions $f_1(\gamma)$, $f_2(\gamma)$, and $f_3(\gamma)$ up to $\mathcal{O}(\lambda^3)$ by matching them order-by-order with the general flow equation~\eqref{ldijnn}. 

An important observation emerging from these calculations is the special case $e_2 = \frac{1}{4}$. In this particular scenario, the differential equations for the functions $f_i(\gamma)$ reduce to the following simple universal form:
\begin{eqnarray}\label{df03i}
 f^{\prime}_i(\gamma)  - (i+1) f_i(\gamma)=0\,.
\end{eqnarray}
The solution to these simplified differential equations~\eqref{df03i} admits the general solution:
\begin{eqnarray}\label{df033i}
 f_i(\gamma)=m_i\, e^{(i+1) \,\gamma}  \,.
\end{eqnarray}
Considering the special case $e_2 = \frac{1}{4}$, we substitute the functions $f_i(\gamma)$ obtained from~\eqref{df033i} into the Lagrangian~\eqref{Gqqn11}. This substitution allows us to explicitly demonstrate that the following root flow equation holds for this deformed Lagrangian:
\begin{equation}\label{eq:root_flow}
    \frac{\partial \mathcal{L}_{GSM}}{\partial \gamma} =\frac{1}{\sqrt{2}} \sqrt{ T_{\mu\nu}T^{\mu\nu} - \frac{1}{2} {T_{\mu}}^{\mu} {T_{\nu}}^{\nu}}=\tau  \, \dot \ell(\tau )
\end{equation}
The explicit form of the root flow equation~\eqref{eq:root_flow} is given by:
\begin{eqnarray}\label{Gqqjjln11}
 \frac{\partial \mathcal{L}_{GSM}}{\partial \gamma}=	\tau  \, \dot \ell(\tau )& =&e^{-\gamma} q_1 + e^{\gamma} q_2 + 2 \,e^{-2 \gamma} m_1 \lambda (- q_1^2 + e^{4 \gamma} q_2^2) \\
    &+& e^{-3 \gamma} \lambda^2 (q_1 + e^{2 \gamma} q_2) \bigl(4 m_1^2 q_1 (3 q_1 -  e^{2 \gamma} q_2) - 3 m_2 (q_1^2 -  e^{4 \gamma} q_2^2)\bigr) \nonumber \\
    &-& e^{-4 \gamma} \lambda^3 \bigl(4 (24 m_1^3 - 12 m_1 m_2 + m_3) q_1^4 + 8 e^{2 \gamma} (14 m_1^3 - 9 m_1 m_2 + m_3) q_1^3 q_2 \nonumber \\
    &-& 8 e^{6 \gamma} (2 m_1^3 - 3 m_1 m_2 + m_3) q_1 q_2^3 -4 e^{8 \gamma} m_3 q_2^4\bigr) \nonumber
\end{eqnarray}
In this section, the standard root flow equation was investigated for the special case $e_2 = \frac{1}{4}$ using the functions $f_i(\gamma)$ defined in Eq.~\eqref{df033i}. The general Lagrangian associated with this case was examined, and it was shown that the perturbation Lagrangian up to $\mathcal{O}(\lambda^3)$ reproduced the developed theories through a suitable choice of the constant coefficients ${m_i}$. By substituting the functions $f_i(\gamma)$ from Eq.~\eqref{df033i} into the Lagrangian given by Eq.~\eqref{Gqqn11}, the following expression was obtained:
\begin{eqnarray}\label{Gqqn11fii}
\mathcal{L}_{GSM} (q_1,q_2)&=&- e^{-\gamma} q_1 + e^{\gamma} q_2 + e^{-2 \gamma} m_1 \lambda (q_1 + e^{2 \gamma} q_2)^2 + \lambda^2 \bigl(-4 e^{-3 \gamma} m_1^2 q_1 (q_1 + e^{2 \gamma} q_2)^2\nonumber \\
    &+& e^{-3 \gamma} m_2 (q_1 + e^{2 \gamma} q_2)^3\bigr) + \lambda^3 \bigl(-12 e^{-4 \gamma} m_1 m_2 q_1 (q_1 + e^{2 \gamma} q_2)^3  \\
    &+& e^{-4 \gamma} m_3 (q_1 + e^{2 \gamma} q_2)^4 + 8 e^{-4 \gamma} m_1^3 q_1 (q_1 + e^{2 \gamma} q_2)^2 (3 q_1 + e^{2 \gamma} q_2)\bigr)\nonumber
\end{eqnarray}
In sec. \eqref{261}, we formulate several integrable theories using the CH-approach, including general Born-Infeld-like, logarithmic, and $q$-deform theories. The extension of all these theories up to order $\lambda^3$ is obtained by selecting an appropriate set of constants $\{m_i\}$ from the Lagrangian~\eqref{Gqqn11fii}. In the following, we briefly outline the CH-functions and Lagrangians associated with these integrable systems.

\subsection{Uniform rescaling}\label{022}
In the study of integrable field theories, a fundamental structure arises in the form of the root flow equation, which governs the deformation of the Lagrangian $\mathcal{L}$ with respect to the flow parameter $\gamma$. For a broad class of known integrable theories, this equation takes the form
\begin{equation}
\frac{\partial \mathcal{L}}{\partial \gamma} = \tau  \dot{\ell}(\tau)\,.
\end{equation}
However, an intriguing simplification occurs in the case of ModMax{-like}  theory in 2D that preserves the conformal invariance, with $T_{\mu}\,^{ \mu}\sim \dot{\ell}(\tau) - \partial_\tau \ell(\tau)=0$. Due to the property $
\ell(\tau) = \tau  \dot{\ell}(\tau),$
the root flow equation reduces to
\begin{equation}\label{eq:ModMaxFlow}
\frac{\partial \mathcal{L}_{\text{ModMax}}}{\partial \gamma} = \ell(\tau).
\end{equation}
This suggests that ModMax theory occupies a special position within the space of integrable deformations.
A natural question then arises:
Is ModMax{-like}  theory the only theory that satisfies the simplified flow equation \eqref{eq:ModMaxFlow}, or does a more general class of such theories exist?

In the continuation of this section, our primary goal is to systematically explore whether there exists an extended family of integrable theories for which the right-hand side of the $\gamma$-flow equation is given directly by the characteristic function $\ell(\tau)$, rather than its derivative-weighted form $\tau  \dot{\ell}(\tau)$. Let us consider the following modified flow equation for a general Lagrangian~\eqref{Gqqn11} corresponding to the CH-function~\eqref{Lexpan}:
\begin{equation}\label{eq:ModMaxFlowGG}
\frac{\partial \mathcal{L}_{GSM}}{\partial \gamma} = \ell(\tau).
\end{equation}
We have previously derived the left-hand side of the flow equation~\eqref{eq:ModMaxFlowGG} in~\eqref{Gqqjjn11}. The right-hand side can be obtained by substituting the auxiliary field value $\tau$ from equation~\eqref{Gqqwn11} into the CH-function~\eqref{Lexpan}. This substitution yields the following expression for the CH-function:
\begin{eqnarray}\label{Gqqjjlln11}
	 \ell(\tau )&=&
    e^{-\gamma} q_1 + e^{\gamma} q_2 + \lambda f_1(\gamma) (-3 e^{-4 \gamma} q_1^2 - 2 e^{-2 \gamma} q_1 q_2 + q_2^2) +\lambda^2 \Bigl(e^{-6 \gamma} f_2(\gamma) \\
    &\times& (-5 q_1 + e^{2 \gamma} q_2)  (q_1 + e^{2 \gamma} q_2)^2 + 4 e^{-7 \gamma} \bigl(f_1(\gamma)\bigr)^2 q_1 (q_1 + e^{2 \gamma} q_2) (5 q_1 + e^{2 \gamma} q_2)\Bigr) \nonumber \\
    &+& \lambda^3 \Bigl( e^{-8 \gamma} f_3(\gamma) (-7 q_1 + e^{2 \gamma} q_2) (q_1 + e^{2 \gamma} q_2)^3 + 12 e^{-9 \gamma} f_1(\gamma) f_2(\gamma) q_1 (q_1 + e^{2 \gamma} q_2)^2\nonumber\\
    &\times& (7 q_1 + e^{2 \gamma} q_2) -  8 e^{-10 \gamma} \bigl(f_1(\gamma)\bigr)^3 q_1 (q_1 + e^{2 \gamma} q_2) (21 q_1^2 + 14 e^{2 \gamma} q_1 q_2 + e^{4 \gamma} q_2^2)\Bigr)\nonumber
\end{eqnarray}
By comparing both sides of the flow equation~\eqref{eq:ModMaxFlowGG}, derived from the CH-function in equation~\eqref{Gqqjjlln11} and the Lagrangian derivative in equation~\eqref{Gqqjjn11} , we obtain, to all orders in $\lambda$, the following differential equation:
\begin{eqnarray}\label{lf01}
 f_i(\gamma)- f^{\prime}_i(\gamma)=0
\end{eqnarray}
Therefore, by solving the preceding equation, we obtain the complete family of solutions for the functions $ f_i(\gamma)$:
\begin{eqnarray}\label{lf011}
 f_i(\gamma)=n_i \,  e^{ \gamma}
\end{eqnarray}
where $n_i$ are constants. By substituting the $ f_i(\gamma)$ from equation \eqref{lf011}, we obtain the following Lagrangian:
\begin{eqnarray}\label{Lfiga}
\mathcal{L}_{GSM}&=&- e^{-\gamma} q_1 + e^{\gamma} q_2 + e^{-3 \gamma} n_1 \lambda (q_1 + e^{2 \gamma} q_2)^2 + \lambda^2 e^{-5 \gamma} \bigl(-4 n_1^2 q_1 (q_1 + e^{2 \gamma} q_2)^2  \\
    &+&  n_2 (q_1 + e^{2 \gamma} q_2)^3\bigr) + \lambda^3 \,e^{-7 \gamma}\bigl(-12  n_1 n_2 q_1 (q_1 + e^{2 \gamma} q_2)^3 +  n3 (q_1 + e^{2 \gamma} q_2)^4  \nonumber \\
    &+& 8 n_1^3 q_1 (q_1 + e^{2 \gamma} q_2)^2 (3 q_1 + e^{2 \gamma} q_2)\bigr)\nonumber
\end{eqnarray}
One can explicitly verify that the Lagrangian~\eqref{Lfiga} expanded to $\mathcal{O}(\lambda^3)$ satisfies the flow equation~\eqref{eq:ModMaxFlowGG}. A comparison of the perturbation theories described in Eqs.~\eqref{Gqqn11fii} and \eqref{Lfiga} reveals that the Lagrangian in \eqref{Lfiga} reduces to that in ~\eqref{Gqqn11fii} under a rescaling transformation of the form $(\lambda \to e^{\gamma}  \lambda)$. This implies that the corresponding flow equations are not independent; specifically, the flow equation \eqref{eq:ModMaxFlowGG} reduces to \eqref{eq:root_flow} under the same rescaling. This relationship suggests that the flow equation given in \eqref{eq:root_flow} is unique.
\subsection{Single trace flow  equations }\label{092}
 In Sec.~\ref{021} and Sec.~\ref{022}, we constructed perturbation theories for the flow equations of the~\eqref{eq:root_flow} and~\eqref{eq:ModMaxFlowGG}. We now consider another special combination that yields the flow equation for the $(\gamma)$ coupling, identified with $\frac{\partial \mathcal{L}}{\partial \gamma}=  \ell (\tau) - \tau \dot \ell (\tau)$. This combination is precisely the energy-momentum trace tensor of the deformed theory:
\begin{eqnarray}\label{7yt}
	\frac{\partial \mathcal{L}}{\partial \gamma}=  \frac{1}{d}  {T_{\mu}}^{\mu}\,.
\end{eqnarray}
Since the ModMax theory does not apply to the above flow equation, we instead consider a CH-function where the two-dimensional seed theory is a principal chiral model (PCM). We therefore introduce a suitable perturbation via this CH-function to generate the flow equation~\eqref{7yt} as follows:
\begin{equation}\label{7yt5}
	\ell (\tau) = \tau  +\sum_{i=1}^\infty  \, \lambda^i g_i(\gamma)  \, \tau^{i+1}\,,
\end{equation}
where the $g_i(\gamma)$ are functions of $\gamma$.
The expansion of the CH-function, including terms up to $\lambda^3$, is given by:
\begin{equation}\label{Lexpangd}
	\ell (\tau) = \tau  +\lambda \,  g_1(\gamma) \,  \tau^{2}  +\lambda^2 \,  g_2(\gamma) \, \tau^{3}  +\lambda^3 \, g_3(\gamma) \,  \tau^{4} +...\,\,.
\end{equation}
Using the $\mathrm{CH}$-function expansion in Eq.~\eqref{Lexpangd} to solve the second equation of Eq.~\eqref{gensol}, we obtain the $\tau$-function to $(\lambda^3)$:
\begin{eqnarray}\label{Lexpang}
	\tau&=& q_1 + q_2 - 4 \lambda g_1(\gamma) q_1 (q_1 + q_2)  + 2 \lambda^2 q_1 (q_1 + q_2) \Bigl( 2 \bigl(g_1(\gamma)\bigr)^2 (7 q_1 + 3 q_2)-3 g_2(\gamma) (q_1 + q_2)\Bigr) \nonumber \\
    &-& 4 \lambda^3 q_1 (q_1 + q_2) \Bigl(2 g_3(\gamma) (q_1 + q_2)^2 - 9 g_1(\gamma) g_2(\gamma) (q_1 + q_2) (3 q_1 + q_2) \\
    &+&  4 \bigl(g_1(\gamma)\bigr)^3 (5 q_1 + q_2) (3 q_1 + 2 q_2)\Bigr)\nonumber\,.
\end{eqnarray}
The deformed Lagrangian to $\mathcal{O}(\lambda^3)$, derived from the CH-function in Eq.~\eqref{Lexpang}, is found by computing $\dot{\ell}(\tau)$ using Eqs.~\eqref{Lexpangd} and \eqref{Lexpang}. It is given by:
\begin{eqnarray}\label{Lag8}
	\mathcal{L} &=& - q_1 + q_2 + \lambda g_1(\gamma) (q_1 + q_2)^2 -  \lambda^2 (q_1 + q_2)^2 \Bigl(4 \bigl(g_1(\gamma)\bigr)^2 q_1 -  g_2(\gamma) (q_1 + q_2)\Bigr)\\
    &+&  \lambda^3 (q_1 + q_2)^2 \Bigl(-12 g_1(\gamma) g_2(\gamma) q_1 (q_1 + q_2) + g_3(\gamma) (q_1 + q_2)^2 + 8 \bigl(g_1(\gamma)\bigr)^3 q_1 (3 q_1 + q_2)\Bigr)  \nonumber  \,.
\end{eqnarray}
The left-hand side of the flow equation~\eqref{7yt} can be obtained by taking the derivative of the Lagrangian~\eqref{Lag8}, as follows:
\begin{eqnarray}\label{DLag}
	\frac{\partial \mathcal{L}}{\partial \gamma} &=& g^{\prime}_1(\gamma) (q_1 + q_2)^2 \lambda + \lambda^2 \bigl(g^{\prime}_2(\gamma) (q_1 + q_2)^3 - 8 g^{\prime}_1(\gamma) q_1 (q_1 + q_2)^2 g_1(\gamma)\bigr) \\
    &+&  \lambda^3 \Bigl(g^{\prime}_3(\gamma) (q_1 + q_2)^4 - 12 g^{\prime}_2(\gamma) q_1 (q_1 + q_2)^3 g_1(\gamma) + 24 g^{\prime}_1(\gamma) q_1 (q_1 + q_2)^2 (3 q_1 + q_2) \bigl(g_1(\gamma)\bigr)^2 \nonumber  \\
    &-&  12 g^{\prime}_1(\gamma) q_1 (q_1 + q_2)^3 g_2(\gamma)\Bigr)  \nonumber  \,.
\end{eqnarray}
The right-hand side of the flow equation~\eqref{7yt} in two dimensions is given by the trace of the energy-momentum tensor of the Lagrangian ~\eqref{Lag8}, as specified by the relation ${T_{\mu}}^{\mu}= 2\, \Big( \ell (\tau) - \tau \dot \ell (\tau)\Big) $:
\begin{eqnarray}\label{DLkagtr}
	\frac{1}{2} {T_{\mu}}^{\mu}&=& - \lambda g_1(\gamma) (q_1 + q_2)^2 +\frac{1}{2} \lambda^2 \Bigl(16 g^2_1(\gamma) q_1 (q_1 + q_2)^2 - 4 g_2(\gamma) (q_1 + q_2)^3\Bigr) \\
    &+&\frac{1}{2} \lambda^3 \Bigl(72 g_1(\gamma) g_2(\gamma) q_1 (q_1 + q_2)^3 - 6 g_3(\gamma) (q_1 + q_2)^4 - 48 g^3_1(\gamma) q_1 (q_1 + q_2)^2 (3 q_1 + q_2)\Bigr) \nonumber\,.
\end{eqnarray}
For the flow equation $\frac{\partial \mathcal{L}}{\partial \gamma}=  \frac{1}{2}  {T_{\mu}}^{\mu}$ to hold, the expressions given in Eqs.~\eqref{DLag} and~\eqref{DLkagtr} must be equivalent. Comparing them order by order in $\lambda$ yields a differential equation of the form $g^{\prime}_i(\gamma)  +  i\,\, g_i(\gamma)=0$. 
The solution to these simplified differential equations admits the general solution:
\begin{eqnarray}\label{dfgg3i}
 g_i(\gamma)=k_i \, \, e^{-\,i \,\gamma}  \,.
\end{eqnarray}
By substituting the  functions $g_i(\gamma)$ from Eq.~\eqref{dfgg3i} into the Lagrangian in Eq.~\eqref{Lag8}, we obtain the following expression:
\begin{eqnarray}\label{Lagam}
	\mathcal{L} &=& - q_1 + q_2 + e^{-\gamma} k_1 \lambda (q_1 + q_2)^2 + e^{-2 \gamma} \lambda^2 (q_1 + q_2)^2 \bigl(-4 k_1^2 q_1 + k_2 (q_1 + q_2)\bigr) \\
    &+& e^{-3 \gamma} \lambda^3 (q_1 + q_2)^2 \bigl(-12 k_1 k_2 q_1 (q_1 + q_2) + k_3 (q_1 + q_2)^2 + 8 k_1^3 q_1 (3 q_1 + q_2)\bigr) \nonumber  \,.
\end{eqnarray}
One can demonstrate explicitly that the Lagrangian~\eqref{Lagam} satisfies the flow equation~\eqref{7yt} in two dimensions. 
We report a single-trace flow equation with respect to the coupling parameter \( \gamma \) within the framework of general perturbation theory. This formulation complements the well-known single-trace flow equation associated with the \( \lambda \) coupling, which takes the form:
$
\tfrac{\partial \mathcal{L}^{(\lambda,\gamma)}}{\partial \lambda} = -\tfrac{1}{d \, \lambda} \, T_\mu{}^\mu,
$
where \( d \) denotes the spacetime dimension and \( T_\mu{}^\mu \) is the trace of the energy–momentum tensor.
By applying the suitable transformation $(\lambda \to e^{-\gamma}  \lambda)$ to the single-trace flow equation $
\tfrac{\partial \mathcal{L}^{(\lambda,\gamma)}}{\partial \lambda} = -\tfrac{1}{d \, \lambda} \, T_\mu{}^\mu,
$ we can systematically derive the corresponding single-trace flow equation~\eqref{7yt} for the \( \gamma \) coupling. { This result demonstrates that no independent, non-trivial $\gamma$-deformation satisfying~\eqref{7yt} exists. It invariably reduces to a $\lambda$-deformation, since the former is obtained by a mere rescaling of the parameter $\lambda$.}
\section{Conclusion and discussion}\label{06kk}
In this work, we presented a general framework for constructing two-dimensional integrable sigma models by employing the Courant-Hilbert (CH) method to solve the universal integrability condition formulated as a partial differential equation in terms of energy-momentum tensor structures. This approach led to a broad class of exactly solvable models parameterized by two couplings, $\gamma$ and $\lambda$, corresponding to marginal and irrelevant deformations, respectively.

We recovered well-known theories, such as the Principal Chiral Model and the two-dimensional ModMax-like theory, as particular CH-function choices, and introduced new integrable models, including generalized Born-Infeld-like, $q$-deformed, and logarithmic theories, that extend the known landscape of integrable field theories. All these models were shown to satisfy a universal root $T\bar{T}$ flow equation, which we proved to be dimension-independent. This allowed us to establish a unifying representation for marginal and irrelevant deformations in both 2D integrable sigma models and 4D duality-invariant electromagnetic theories.
\subsection*{Non-conformal  family of  ModMax theories}\label{84122}

We introduce the {\it non-conformal family of ModMax theories}, a new class of integrable models defined by root flow equations distinct from those previously studied. ModMax theories emerge as a special case, where conformal symmetry is enforced $\ell=\tau\dot{\ell}$, reducing the general root flow equation  
$
\frac{\partial \mathcal{L}}{\partial \gamma}=\tau \dot{\ell}(\tau)
$ 
to  
$
\frac{\partial \mathcal{L}}{\partial \gamma}=\ell(\tau).
$
We investigate whether ModMax is unique in satisfying this stronger condition, or if a broader class of analytic, integrable theories exists whose $\gamma$-flows are governed directly by the CH-function $\ell(\tau)$, providing explicit two-dimensional examples.
We begin by assuming that the integrable Lagrangians applicable to the flow equation of the $
\frac{\partial \mathcal{L}}{\partial \gamma}=\ell(\tau).
$ exhibit root-type flow equations. However, the coefficients between the two structures, $X= \tfrac{T_{\mu\nu}T^{\mu\nu}}{d}$ and $Y= \tfrac{{T_{\mu}}^{ \mu}\,{T_{\nu}}^{ \nu}}{d^2} $, differ. Consequently, this type of Lagrangian must apply to the root-type flow equation as follows:
\begin{equation}\label{lM28dijnn}	 \frac{\partial \mathcal{L}}{\partial \gamma}=\sqrt{X-n\,Y} = \ell (\tau)\,.
\end{equation}
where $n$ is a constant. We can form a differential equation by substituting structures $X$ and $Y$ from ~\eqref{Tnunudi} into ~\eqref{lM28dijnn}. This differential equation is as follows:
\begin{equation}\label{PDEell}	
	-n\,{\ell (\tau)}^2+2 (n-1) \tau \ell (\tau) \dot{\ell}(\tau)-(n-2)\,\tau^2 {\dot{\ell}(\tau)}^2 =0 .
\end{equation}
The above equation reduces to the two limiting cases $n \to 1$ and $n \to \infty$ under the condition ${\ell (\tau)} -\tau {\dot{\ell}(\tau)}=0 $. In these special limits, the solution to the differential equation is given by the CH-function of the ModMax theory in the form $\ell (\tau) =a\,\, \tau$, where $a$ is $e^{\gamma}$. However, there is another solution to the differential equation ~\eqref{PDEell}, known as the non-trivial solution. This solution is related to the CH-function of a new family of theories, which we refer to as the non-conformal family of ModMax theories. By solving the differential equation ~\eqref{PDEell}, the CH-function of these theorems can be derived as follows:
\begin{equation}\label{PDEellso}
	\ell (\tau) =a\,\, \tau^{\mathbf{m}}\,,
\end{equation}
where are $\mathbf{m}={\frac{n}{n-2}}$ and $a= \lambda^{\mathbf{m}-1} e^{\gamma}$. Since the CH-function ~\eqref{PDEellso} lacks a perturbative expansion, these theories do not possess a limiting state. In other words, there is no specific limit to which these theories can be reduced to ModMax theories. In appendix, we examine theories with a CH-function of  ~\eqref{PDEellso} and explicitly derive two-dimensional integrable theories for $\mathbf{m}={\frac{1}{2}}$ and $\mathbf{m}={\frac{3}{2}}$. These theories satisfy the root-type flow equation ~\eqref{lM28dijnn} by $n=-2$ and $n=6$, respectively. 
We can study two explicit examples of integral theories that satisfy the flow equation $\frac{\partial \mathcal{L}}{\partial \gamma}= \ell (\tau)\,$. This group of integral theories forms a large class of "{\it{non-conformal family of ModMax theories}}". This type of integrable theory can be generated to higher orders in $\lambda$ as a closed form of the appropriate CH-function. In this case, the flow equation~\eqref{lM28dijnn} is still preserved for these theories. We will investigate this issue in future work and provide illustrative examples. 
One important observation regarding the closely related family of ModMax-type theories is that these theories lack conformal symmetry; in other words, $T_{\mu}\,^{ \mu} \neq 0$. Furthermore, they do not reduce to PCM in the limit $\gamma \to 0$. Nevertheless, these theories are integrable and possess a flow equation analogous to that of the MadMax theory with $\gamma$ coupling.

Moreover, we identified a non-conformal family of ModMax-type theories governed by alternative root-type flow equations, revealing the existence of nonperturbative integrable models beyond the standard ModMax structure. These examples illustrate the effectiveness of the CH approach in systematically generating new solvable deformations while maintaining integrability.

\subsection*{Frame transformations and the rescaling of root-$T\overline{T}$ deformations}
In string theory, the tension $T_p$ of a D$p$-brane in the string frame plays a fundamental role in determining the dynamics of extended objects. This tension is given by~\cite{Garousi_2017}
\begin{equation}
    T_p = \frac{1}{g_s (2\pi)^p (\alpha')^{(p+1)/2}},
    \label{eq:tension}
\end{equation}
where $g_s \equiv e^{\phi_0}$ denotes the closed string coupling constant, and $\alpha'$ sets the fundamental length scale in string theory. The dependence on $g_s$ highlights that D-brane tensions are non-perturbative quantities in string theory, scaling inversely with the string coupling~\cite{Tong:2009np}.

The string-frame metric $g^S_{\mu\nu}$ is related to the Einstein-frame metric $g^E_{\mu\nu}$ via the dilaton field $\phi_0$:
\begin{equation}
    g^S_{\mu\nu} = e^{\phi_0/2} \, g^E_{\mu\nu}.
    \label{eq:frame_transform}
\end{equation}
The corresponding transformation of the D$p$-brane tension between frames takes the form
\begin{equation}
    T_p^E = T_p^S \, e^{-\frac{(p-3)}{4}\phi_0}.
\end{equation}

For the D1-brane ($p=1$), one obtains
\begin{equation}
    T_1^E = T_1^S \, e^{-\frac{(1-3)}{4}\phi_0} 
           = T_1^S \, e^{\frac{1}{2}\phi_0} 
           = T_1^S \, g_s^{1/2}.
    \label{eq:consistent_transform}
\end{equation}
For the D3-brane ($p=3$), the relation reduces to
\begin{equation}
    T_3^E = T_3^S,
    \label{eq:d3}
\end{equation}
indicating that the D3-brane tension is frame-independent. This invariance is closely related to the S-duality symmetry of type IIB string theory~\cite{Garousi:2011vs,PhysRevD.88.026008,BabaeiVelni:2016qea}.

If one interprets $T_p = 1/\lambda$, it may initially appear that the rescaling of the coupling $\lambda$ mimics the transformations between the Einstein and string frames. However, since the D3-brane tension is invariant under such frame changes, the rescaling of $\lambda$ cannot be directly attributed to frame transformations. Moreover, the dilaton is a scalar field in string theory with intrinsic dynamics, fundamentally distinguishing it from the $\gamma$ coupling, which is dimensionless. It is conceivable that developing a framework in which the $\gamma$ coupling acquires dynamics could lead to the discovery of a novel field in string theory. Such a field might generate new dynamical structures and naturally give rise to marginal flow equations in the context of string theory.

In our forthcoming work, we aim to elaborate on a new approach to integrable theories based on the new auxiliary field formulation introduced in~\cite{Lechner:2022qhb,Russo:2025fuc}, which we have recently summarized in~\cite{Babaei-Aghbolagh:2025uoz}. This framework unifies deformation flows through a two-parameter space governed by irrelevant ($\lambda$) and marginal ($\gamma$) couplings, offering a systematic extension of the Courant-Hilbert method. A key feature of the construction is the emergence of a root-$T\bar{T}$ flow condition, which serves as a universal integrability constraint across both two-dimensional and four-dimensional sectors.
An intriguing direction builds upon the recent work of Russo and Townsend~\cite{Russo:2025fuc}, who developed causal formulations of self-dual nonlinear electrodynamics and chiral 2-form theories~\cite{Russo:2025wph}. Their Hamiltonian-based parameterization ensures compliance with the Dominant and Strong energy conditions, providing a robust criterion for causality in nonlinear regimes. We observe that their stress-energy tensor analysis aligns with the convexity conditions derived from our deformation potential, suggesting a deeper geometric structure underlying causal integrable flows.

This synergy opens the possibility of embedding causal self-dual electrodynamics within a broader integrable landscape, potentially leading to new classes of deformation-driven models that respect both duality and causality.

\section*{Acknowledgments}
We are grateful to Jue Hou, J. G. Russo, and Dmitri Sorokin for their interest in this work and the fruitful discussions that followed. 
H.B.-A. would like to express my sincere gratitude to Karapet Mkrtchyan, Neil Lambert and Alessandro Tomasiello for the highly productive discussions during the "Journey through Modern Explorations in QFT and beyond - 2 $\otimes$ 25" conference in Yerevan (August 3-13).
The work of H.B.-A. was conducted as part of the PostDoc Program on {\it Exploring TT-bar Deformations: Quantum Field Theory and Applications}, sponsored by Ningbo University. This research was partly supported by NSFC Grant No.11735001, 12275004, 12475053, 12235016.

\bibliographystyle{JHEP}

\providecommand{\href}[2]{#2}\begingroup\raggedright\endgroup

\end{document}